\documentclass[a4paper,12pt]{article}
\usepackage[a4paper,left=2.5cm, right=2.5cm, top=2.5cm, bottom=2.5cm]{geometry}
\usepackage[latin1]{inputenc}   	
\usepackage[T1,TS1]{fontenc}        

\usepackage[colorinlistoftodos, textwidth=6.5cm, shadow]{todonotes}
\usepackage[doublespacing]{setspace} 
\usepackage{color}              
\usepackage{amsmath,amsthm,amsfonts,amssymb,amscd,mathrsfs}

\usepackage{bm}	
\usepackage{bbm}
\usepackage[ngerman, english]{babel}     %
\usepackage{ae}                 %
\usepackage{graphicx}           
\usepackage{epstopdf}
\usepackage{braket}
\usepackage{longtable}          
\usepackage{array}
\usepackage[inline]{enumitem}
\usepackage{multirow}           
\usepackage{url}
\usepackage{pdfpages}
\usepackage{epsfig}
\usepackage{fancybox}
\usepackage[font={bf},labelfont={bf}, hypcap=false,format=hang]{caption}
\usepackage{subcaption}
\usepackage{bold-extra}
\usepackage{titlesec}
\usepackage{xcolor}
\usepackage{fancyvrb}
\usepackage{arydshln}
\usepackage{lscape}
\usepackage{import}
\usepackage[titletoc]{appendix}
\usepackage{todonotes}
\usepackage[babel]{csquotes}
\usepackage{datetime}
\newdateformat{deformat}{\THEDAY{.} \monthname[\THEMONTH] \THEYEAR}
\definecolor{tuebingendarkred}{RGB}{165,30,55}
\usepackage{alphalph}
\usepackage{pifont}
\usepackage{microtype}
\usepackage{placeins}
\usepackage{MnSymbol}
\usepackage{tikz}
\usetikzlibrary{arrows,snakes,shapes,automata,backgrounds,petri,calc,matrix,decorations.markings,decorations.pathreplacing,patterns,arrows.meta}
\usepackage{pgf}
\usepackage{adjustbox}
\usepackage{booktabs}
\usepackage{multirow}

\makeatletter

\renewcommand\tableofcontents{%
  \null\hfill\textbf{\Large\contentsname}\hfill\null\par
  \@mkboth{\MakeUppercase\contentsname}{\MakeUppercase\contentsname}%
  \@starttoc{toc}%
}
\renewcommand\listoffigures{%
  \null\textbf{\large\listfigurename}\hfill\null\par
  \@mkboth{\MakeUppercase\listfigurename}{\MakeUppercase\listfigurename}%
  \@starttoc{lof}%
}
\renewcommand\listoftables{%
  \null\textbf{\large\listtablename}\hfill\null\par
  \@mkboth{\MakeUppercase\listtablename}{\MakeUppercase\listtablename}%
  \@starttoc{lot}%
}
\makeatother

\makeatletter
	\renewcommand\appendix{
		\addtocontents{toc}{\null\noindent\large \bfseries Appendices\normalsize\par} 
	\renewcommand\section{
			\newpage
			\thispagestyle{plain}
			\suppressfloats[t]
			\@afterindentfalse
			\secdef\Appendix\sAppendix
		}
		\setcounter{section}{0}%
		\setcounter{figure}{0}%
		\setcounter{equation}{0}%
		\renewcommand\thesection{\textsc{\Alph{section}}}%
		\renewcommand\theequation{\Alph{section}\arabic{equation}}%
		\renewcommand\thefigure{\Alph{section}\arabic{figure}}
	}
	
	\newcommand\Appendix[2][?]{
		\refstepcounter{section}
		\addcontentsline{toc}{section}%
		{\protect\numberline
		{\thesection}#1} {\centering \large \scshape \appendixname\,
		 \thesection \par#2\par}
		 \sectionmark{#1}
		 \@afterheading
		 \addvspace{\baselineskip}
	}
	\newcommand\sAppendix[1]{
		{\raggedleft\large\bfseries\appendixname\
		 \thesection\par \centering#1\par}
		 \@afterheading
		 \addvspace{\baselineskip}
	}
\makeatother

\RecustomVerbatimCommand{\VerbatimInput}{VerbatimInput}%
{fontsize=\footnotesize,
 frame=lines,  
 framesep=2em, 
 rulecolor=\color{Gray},
 label=\fbox{\color{Black}data.txt},
 labelposition=topline,
 commandchars=\|\(\), 
 commentchar=*        
}

\newtheoremstyle{LOBstyle}
  {2\topsep} 
  {\topsep} 
  {\itshape} 
  {} 
  {\bfseries} 
  {.} 
  {.5em} 
  {} 

\theoremstyle{LOBstyle}
\newtheorem{LOB}{Rule}

\theoremstyle{remark}
\newtheorem*{rem}{Remark}

\theoremstyle{plain}

\newcommand{\norm}[1]{\left\Vert#1\right\Vert}

\newcommand{\seq}[1]{\left<#1\right>}

\DeclareMathOperator*{\E}{\mathbb{E}} 	

\DeclareMathOperator*{\diag}{diag}		
\DeclareMathOperator*{\rank}{rank}		

\makeatletter
\def\munderbar#1{\underline{\sbox\tw@{$#1$}\dp\tw@\z@\box\tw@}}
\let\@@pmod\pmod
\DeclareRobustCommand{\pmod}{\@ifstar\@pmods\@@pmod}
\def\@pmods#1{\mkern4mu({\operator@font mod}\mkern 6mu#1)}
\makeatother
\makeatletter 
\def\leqn{\tagsleft@true} 
\def\reqn{\tagsleft@false} 
\def\fleq{\@fleqntrue \let\mathindent\@mathmargin \@mathmargin=\leftmargini} 
\def\cneq{\@fleqnfalse} 
\makeatother 

\newcolumntype{C}[1]{>{\centering\arraybackslash}m{#1}}
\newcolumntype{R}[1]{>{\raggedleft\arraybackslash}m{#1}}  

\usepackage{simpler-wick}

\usepackage{forest}
\usepackage[colorlinks=true, linkcolor=blue, citecolor=blue]{hyperref}
\usepackage{cleveref}

\addto\extrasenglish{

}

\usepackage{booktabs}
\usepackage{environ}
\usepackage{expl3}
\ExplSyntaxOn
\seq_new:N \l_DT_contents_seq
\seq_new:N \l_DT_contents_A_seq
\tl_new:N \l_DT_header_tl
\cs_generate_variant:Nn \seq_set_split:Nnn { NnV }
\NewEnviron{dbltbl}[2]
  {
    \seq_set_split:NnV \l_DT_contents_seq { \\ } \BODY

    \seq_pop_left:NN \l_DT_contents_seq \l_DT_header_tl

    \seq_clear:N \l_DT_contents_A_seq
    \prg_replicate:nn
      { \int_div_truncate:nn { \seq_count:N \l_DT_contents_seq } {2} }
      {
        \seq_pop:NN \l_DT_contents_seq \l_tmpa_tl
        \seq_put_right:NV \l_DT_contents_A_seq \l_tmpa_tl
      }

    \begin{tabular}{#1p{#2}#1}
      \toprule
      \l_DT_header_tl && \l_DT_header_tl \\
      \midrule
      \seq_mapthread_function:NNN
        \l_DT_contents_A_seq
        \l_DT_contents_seq
        \DT_one_row:nn
      \bottomrule
    \end{tabular}
  }
\cs_new:Npn \DT_one_row:nn #1#2 { #1 && #2 \\ }
\ExplSyntaxOff

\usepackage{soul}

\usepackage{fontawesome}
\titlespacing*{\section}{0pt}{4pt}{10pt}

\usepackage[longnamesfirst]{natbib}

\begin{document}

\author{
Johannes Bleher\footnote{Corresponding author. Computational Science Hub (CSH) University of Hohenheim Fruwirthstr. 49,
70599 Stuttgart, Phone: +49 711 459 23431. Email: \texttt{johannes.bleher@uni-hohenheim.de} }\\ \vspace*{-0.3cm} \small {\em University of Hohenheim}, \and
Michael Bleher\footnote{Institute for Mathematics, Heidelberg University. Mathematikon, Im Neuenheimer Feld 205, 69120 Heidelberg. Email:
\texttt{mbleher@mathi.uni-heidelberg.de}}\\ \vspace*{-0.3cm} \small {\em Heidelberg University}
}


\title{An Algebraic Framework for the Modeling of Limit Order Books}
\date{\today}

\singlespace

{\let\newpage\relax\maketitle}

\thispagestyle{empty}
\vspace*{-0.5cm}
\begin{abstract}
  tbd

\end{abstract}
\vspace{0.5cm}

\begin{tabular}{p{0.15\textwidth} p{0.7\textwidth}}
	  \textit{Key words:} &Limit Order Book, Master Equation, Continuous Markov Process, High Frequency, Market Microstructure\\
	  \\
	\textit{JEL:} & C58, D43, G12 \\
\end{tabular}

\begin{abstract}
 Introducing an algebraic framework for modeling limit order books (LOBs) with tools from physics and stochastic processes, our proposed framework captures the creation and annihilation of orders, order matching, and the time evolution of the LOB state. It also enables compositional settings, accommodating the interaction of heterogeneous traders and different market structures.
 We employ Dirac notation and generalized generating functions to describe the state space and dynamics of LOBs.
 The utility of this framework is shown through simulations of simplified market scenarios, illustrating how variations in trader behavior impact key market observables such as spread, return volatility, and liquidity.
 The algebraic representation allows for exact simulations using the Gillespie algorithm, providing a robust tool for exploring the implications of market design and policy changes on LOB dynamics.
 Future research can expand this framework to incorporate more complex order types, adaptive event rates, and multi-asset trading environments, offering deeper insights into market microstructure and trader behavior and estimation of key drivers for market microstructure dynamics.
\end{abstract}

\newpage

\pagestyle{plain}
\pagenumbering{Roman}
\singlespacing

\newpage

\pagenumbering{arabic}
\setcounter{page}{1}
\onehalfspacing

\section{Introduction}
In the quest to achieve a deeper and more precise understanding of the complex dynamics of limit order books (LOBs), a variety of modeling approaches have been added to the literature. To address these complexities, innovative mathematical frameworks have been explored over the years. These tools enrich the analytical landscape and promise to enhance the precision of simulations and forecasts in financial markets. As noted by \cite{HongPage01}, they may expand the set of heuristics available to analyze the order book.

\cite{GouldPWMFH13} provides a comprehensive overview of the literature on market microstructure. Foundational work by \cite{Garman1976}, \cite{Roll1984}, \cite{Kyle85}, \cite{AdmatiP88}, \cite{BiaisHS95}, \cite{Parlour98}, and \cite{SmithFGK03} has significantly contributed to this field.

However, models that incorporate the decisions of market participants and provide nuanced dynamics of all LOB observables are rare. Notable exceptions include \cite{SmithFGK03} and \cite{ContST10}. Often, models focus on the strategic aspects of traders and derive approximating stochastic characteristics of LOB observables from these strategic models. The impact of order flow has been a primary focus in research, as established by the works of \cite{LilloFM03}, \cite{FarmerGLSS04}, and more recently by \cite{ZarinelliTFL15}, \cite{CuratoGL17}, \cite{TarantoBBLT18a}, \cite{TarantoBBLT18b}, \cite{TothPLF15}, and \cite{BonartL18}.

In this paper, we contribute to the literature by offering a new perspective on LOBs through an algebraic framework that describes their dynamics and allows for compositional settings involving heterogeneous traders and different marketplaces. This paper builds on the algebraic notations discussed in \cite{BleherBleherD2021} and \cite{Bleher21}.

We utilize the well-established Dirac notation from physics and generalized generating functions, which are cornerstones in formal power series and combinatorics. Generating functions offer elegant solutions to complex problems involving differential equations and recurrence relations. In the context of LOBs, they facilitate a compact representation of infinite series and provide a functional form that encapsulates all moments of a given random variable, capturing both typical and atypical behaviors within the market.

Furthermore, Dirac notation, a staple in quantum mechanics, allows for an abstract and powerful description of state spaces. By adopting this notation in financial markets, we can describe the state of an LOB clearly and flexibly. Dirac notation is not just a theoretical embellishment but a practical tool that simplifies the formulation and manipulation of complex models.

This paper aims to connect these mathematical concepts with the practical challenges of modeling LOBs. We explore how generating functions serve as basis functions within Dirac notation, enhancing the mathematical tractability of models and ensuring they faithfully represent underlying market dynamics. By redefining LOB dynamics into an algebraic structure that leverages these tools, we provide a robust framework for handling statistical complexities and incorporating strategic behaviors of market actors.

In the following sections, we will summarize the concept of generating functions in \Cref{sec:background} and delineate their role as basis functions in Dirac notation. We will demonstrate how selecting appropriate basis functions influences the derived functionals and impacts the characterization of the LOB's state space. Through this exploration, we will provide details in \Cref{sec:dirac-representation} on the rules of a typical limit order book and explain how they translate into an algebraic representation of creation and annihilation operators.

We argue that reformulating the dynamics of the LOB into an algebra that uses Dirac notation and the notion of generating functions can be a helpful tool for simulating, forecasting, and modeling order book dynamics.

\begin{figure}
	\centering
	\caption{Overview on LOB states}
	\label{fig:LOBStates}
	\begin{minipage}{0.9\linewidth}
		The figures provide a graphical illustration of LOB configurations. Figure~\ref{subfig:OneState} depicts a single LOB state. Ask orders are displayed in green on the right and bid orders in yellow on the left, with prices increasing along the vertical axis. Figure~\ref{subfig:TimeEvolution} depicts the stochastic time evolution of a LOB state as a branching process of varying probabilities, depending on the likeliness of the corresponding order arrival.  Figure~\ref{subfig:LOBOneArrivals} shows how arrivals from different agents vary over different areas on the price line and aggregate to an overall distribution of arrival rates over price levels. Finally, Figure~\ref{subfig:multiState} displays such an arrival rate aggregation for traders that participate in several order books at the same time to emphasize the modularity of our approach.
	\end{minipage}

	\begin{tabular}{lr}
	\begin{subfigure}{0.4\linewidth}
	\caption{One LOB state}
	\label{subfig:OneState}
	 \includegraphics[width=\linewidth]{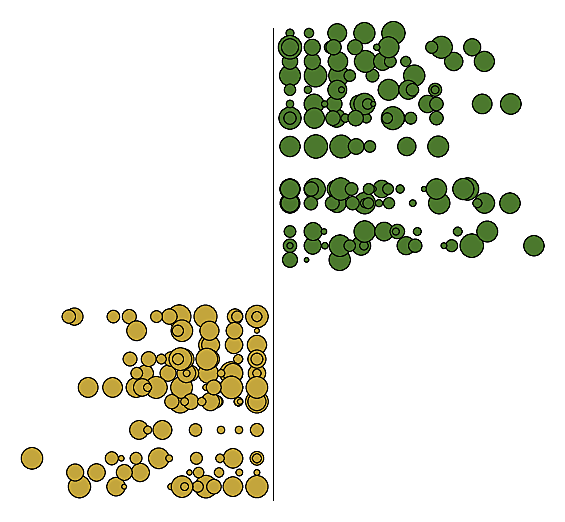}
	\end{subfigure} & \multirow{3}{20cm}{%
	\begin{subfigure}{0.4\linewidth}
	\caption{Time Evolution of a LOB state}
	\label{subfig:TimeEvolution}
	\vspace{1cm}
	 \includegraphics[width=\linewidth]{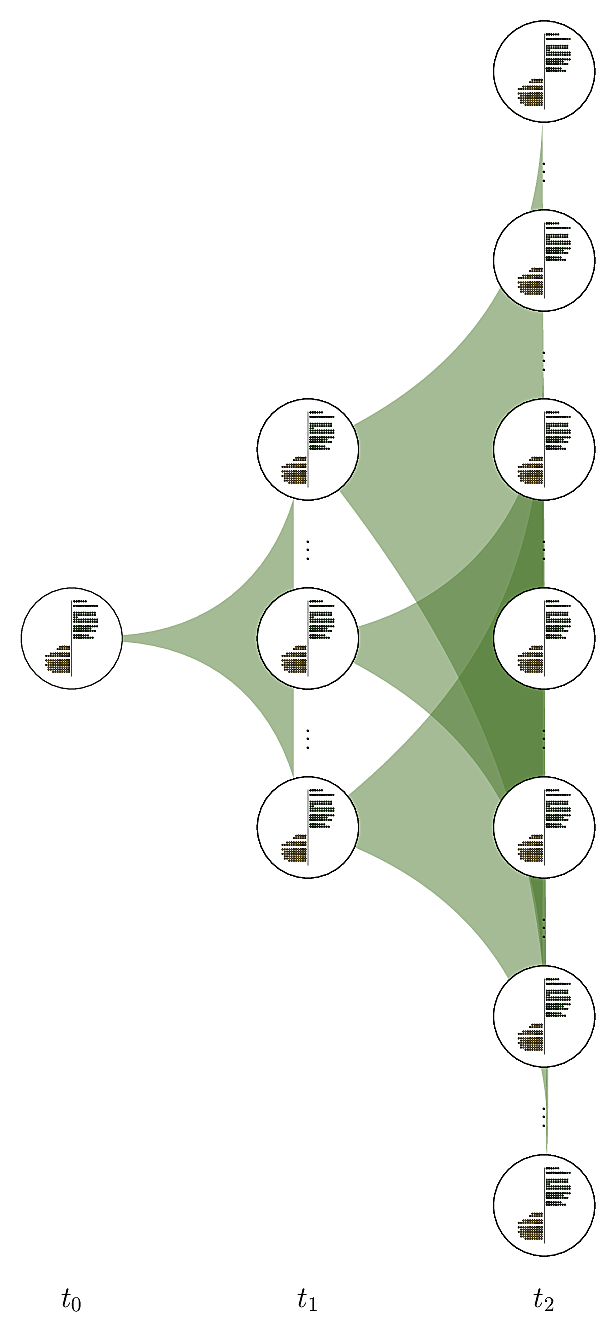}
	\end{subfigure}
	}\\
	\begin{subfigure}{0.4\linewidth}
	\caption{LOB Arrival rates}
	\label{subfig:LOBOneArrivals}
	 \includegraphics[width=\linewidth]{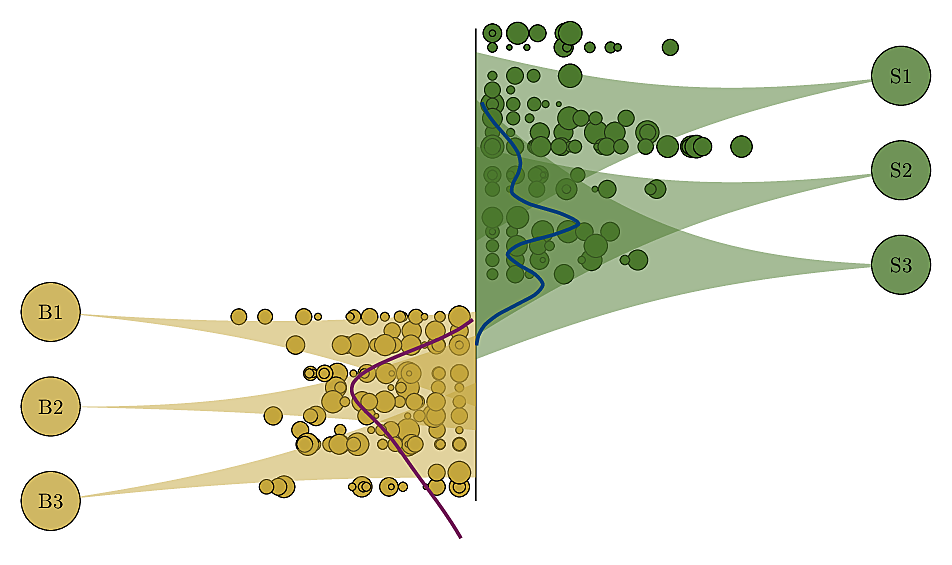}
	\end{subfigure}&\\
	\begin{subfigure}{0.4\linewidth}
	\caption{Multi-LOB state}
	\label{subfig:multiState}
	 \includegraphics[width=\linewidth]{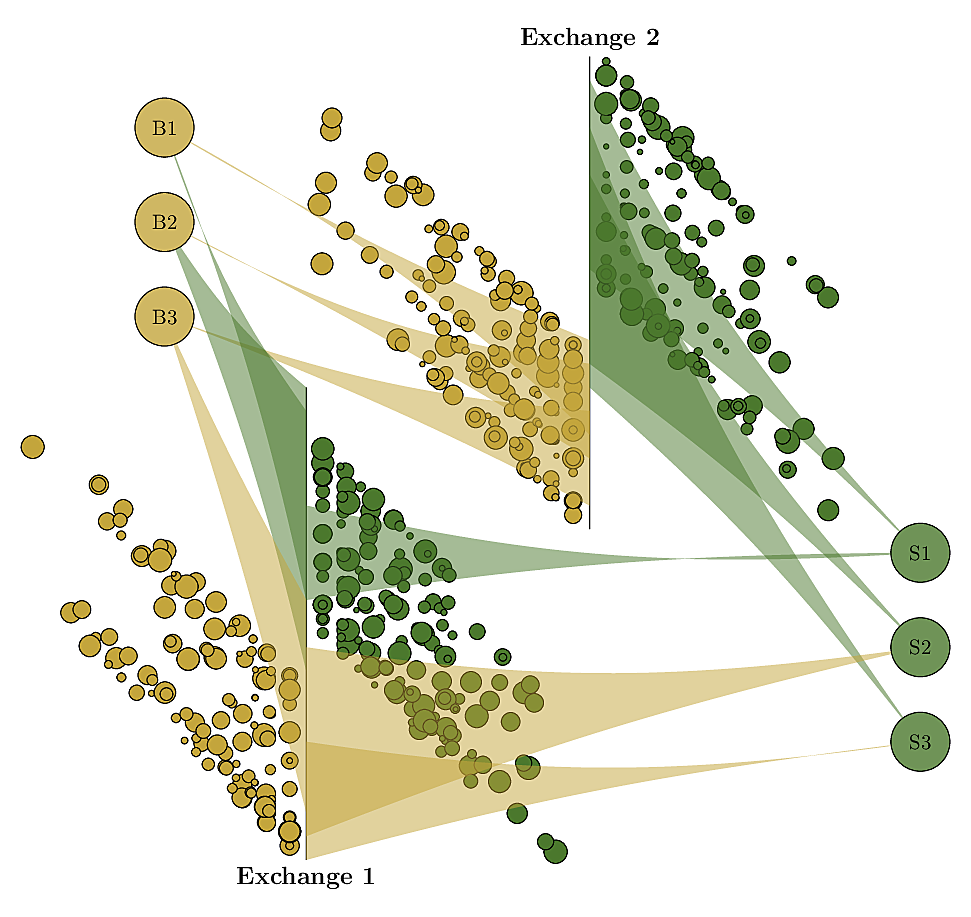}
	\end{subfigure}&\\
	\end{tabular}
\end{figure}

\section{Background}
\label{sec:background}
The description in this section closly follows \cite{BleherBleherD2021} and \cite{Bleher21}.

\subsection{The Limit Order Book}

The limit order book is the place where traders' orders meet.
Their orders carry information about the willingness of the issuing trader to accept a certain price in exchange for the chosen number of instruments.

The price level at which two orders are matched is called \emph{the reference price}.
To the LOB, buy (ask) orders and sell (bid) orders are submitted.
At any point in time, the exchange keeps track of all orders within the LOB.
Depending on the trading mode this orders are matched.
During \textit{continuous trading} incoming orders are directly matched with the already residing orders in the book.
In an \textit{auction}, the exchange usually collects all orders in the LOB until a pre-defined point in time at which the exchange then matches the orders.

Aside from the market side, the location within the LOB is defined by three key components: the limit price, the number of securities as wells as the time when the order arrives in the LOB. All three are determined by the trader -- upto latency considerations that may affect the point in time the order arrives.

If traders require immediacy, they rely on \emph{market orders} which can be thought to have an infinite (bid order) or zero (ask order) limit price depending on the market side on which they were issued.
During continuous trading, market orders are matched immediately and, normally, do not reside in the order book.
They enter the LOB at the best price level of all limit orders currently residing in the LOB on the opposite market side. Since they require immediate matching, the price impact for the issuing trader can be large.

The majority of orders, however, are designed to remain in the book for some time at their specified limit price level \cite[see, e.g.,][]{FreyS2009}.
These are called \emph{limit orders}.
Limit orders have a well defined location in the price dimension.
If their designated price location is behind the best price level of all limit orders which currently reside in the LOB on the other side of the market, they are
matched (partially) before they can reach their designated limit price level.
The smallest populated price level on the sell side of the market is the \emph{best ask} and the highest price level on the bid side is the \emph{best bid}.
We will refer to one or the other as the \emph{best quote}.

There are generally further order types that are only submitted to the market if certain conditions are met.
For example, \emph{stop orders} are inserted in the LOB once the reference price -- usually the price of the last trade -- hits a certain threshold price.
As soon as such an order enters the LOB, it is effectively equivalent to a market or a limit order.
Hence, such conditional orders can be perceived as more sophisticated versions of limit or market orders and can in principle be incorporated in the framework presented below.
For the purpose of this paper, we restrict our considerations to plain limit orders. 
Note again that we consider market orders to be a special kind of limit order with a zero limit price (ask-order) or an infinite limit price (bid order).

\subsection{Probability Generating Functions}

A \emph{generating function} is a representation of a sequence $(a_k)_{k\geq 0}$ as the coefficients of a formal power series $G(a_k ; z) = \sum_{k=0}^\infty a_k z^k$.
Denote the probability distribution of a discrete random variable $X$ by $p_k = \operatorname{Prob}(X=k)$, $k\geq 0$.
The \emph{probability generating function} (PGF) of $X$ is the generating function of the sequence $(p_k)$:
\begin{align}
	G_X(z) := G(p_k; z) = \sum_{k\geq 0} p_k z^k \ .
\end{align}
Note that $G_X(z) = \mathbb{E}[z^X]$, which is sometimes used as a definition of the PGF of $X$ that directly extends to continuous random variables.
In that case it is also called the $z$-transform of the probability mass function of $X$.

The PGF of a random variable $G_X(z) = \sum_{k\geq 0} p_k z^k$ represents what is known as a mixed state: a probabilistically weighted linear combination of the possible outcomes $k \in \mathbb{Z}$ of the discrete random variable $X$.
The evolution of a dynamical stochastic system is captured by a time-dependent mixed state of the form
\begin{align}
 G_X(z, t; k_0, t_0) = \sum p(k, t | k_0, t_0) z^k .
\end{align}
Many stochastic processes can be described by a linear partial differential equation (PDE) for $G_X(z,t)$, see for example \cite{Weber2017} who give a good overview on PDEs for generating functions for different processes.
In this article we investigate the evolution of the limit order book in terms of a time-dependent PGF under the assumption that the dynamics follows a master equation.

\begin{rem}
The PGF is directly related to various other generating functions that are commonly used in statistics and finance.
For example, $G_X(z) = \mathbb{E}[z^X]$ yields the characteristic function of a random variable by setting $z = \exp(s)$ where $s \in \mathbb{C}$: ${CF}_X(s) = \mathbb{E}[e^{sX}]$.
By evaluating ${CF}_X(s)$ on the real line, one recovers the moment generating function ${MGF}_X(t) = \mathbb{E}[e^{tX}]$, $t \in \mathbb{R}$.
From the MGF on can then construct the cumulant generating function by taking the logarithm: $CGF(t) = \ln \mathbb{E}[e^{tX}]$.
Moreover, $G_X(z^\ast +1)$ yields the factorial moment generating function.
\cite{Hoyt72} provides a concise overview over the relations between various generating functions.
\end{rem}

\subsection{Dirac Notation}

Mixed states are common in theoretical physics, in particular quantum mechanics, where they describe a fundamental property of nature.
In this context one often uses the more versatile \emph{Dirac notation}, where the fundamental, or pure, states of a discrete system are represented by \emph{ket}-vectors $\ket{k}$ and a general mixed state is given by a linear combination
\begin{align}
	\ket{X} = \sum p_k \ket{k} \ .
\end{align}
The Dirac notation is particularly advantageous when states arise from a vacuum state $\ket{0}$ by successively adding or removing "excitations".
Creating an additional "excitation" is encoded by the action of a \textit{creation operator} ${a^+ \ket{k} = \ket{k+1}}$ and removing one of the excitations is described by an \textit{annihilation operator} ${\frac{1}{k!} a^- \ket{k} = \ket{k-1}}$.

The creation and annihilation properties of the two operators $a^+$ and $a^-$ are equivalent to the \emph{canonical commutation relation} ${[a^-, a^+] := a^- a^+ - a^+ a^- = 1}$ together with the vaccum condition ${a^- \ket{0} = 0}$:
\begin{align*}
	a^- a^+ \ket{0}
	= (1 + a^+ \underbrace{a^-) \ket{0}}_{=0} 
	= \ket{0} \ .
\end{align*}

The Dirac notation is completely equivalent to the PGF by identifying its pure states $\ket{k}$ with $z^k$.
The mixed state of a random variable $X$ is then equivalently described by
\begin{align}
	\ket{X} = \sum p_k \ket{k} \equiv \sum p_k z^k = G_X(z) \ .
\end{align}
Moreover, under this identification creation and annihilation operators translate to derivatives and multiplications in the variable $z$ (cf. \autoref{sec:pgf-as-polynomial-representation}).

The description of "excitations" in quantum mechanics nicely matches the situation of the order book, which is described by a collection of open ask and bid orders that reside in an otherwise empty order book.
Just like "excitations", order submissions can be represented by a sequence of creation operators acting successively on the order book.
As will be explained in \autoref{sec:dirac-representation}, the order book is described by a large number of distinct types of "excitations".
As a result, the description in terms of a PGF becomes unwieldy, which is ultimately the reason we borrow the Dirac notation from physics.

\begin{rem}
An advantage of generating functions over the Dirac notation appears when the formal power series converges to a simple analytic function.
For example, the PGF of a Poisson-distributed random variable $X \sim \operatorname{Pois}(\lambda)$ is given by ${G_X(z) = \sum_{k\geq 0} \frac{\lambda^k}{k!} e^{-\lambda} z^k = e^{\lambda(z-1)}}$.
\end{rem}

\section{Algebraic Framework for the Limit Order Book}
\label{sec:dirac-representation}

\begin{figure}
	\caption{Rules of the Limit Order Book}
	\label{fig:rules}
	\begin{minipage}{0.9\textwidth}
		The rules underlying a typical limit order book can be translated into an algebraic framework in terms of creation and annihilation operators that act on the current order book state.
		\autoref{subfig:submission} depicts bid and ask order submission into an otherwise empty order book, as prescribed by \autoref{rule:creation}.
		\autoref{subfig:cancellation} shows the order cancellation mechanism \autoref{rule:cancellation}.
		\autoref{subfig:priority} depcits how a newly submitted walks the book until it reaches its limit price and the corresponding commutation of order operators as given by \autoref{rule:priority}. 
		Finally, \autoref{subfig:matching} gives a graphical illustration of partial order matching determined by \autoref{rule:matching}.
	\end{minipage}

	\vspace{1em}

	\begin{subfigure}{0.28\textwidth}
		\caption{Order Submission}
		\label{subfig:submission}
		\centering
		\includegraphics[width=12cm, height=4.6cm, keepaspectratio]{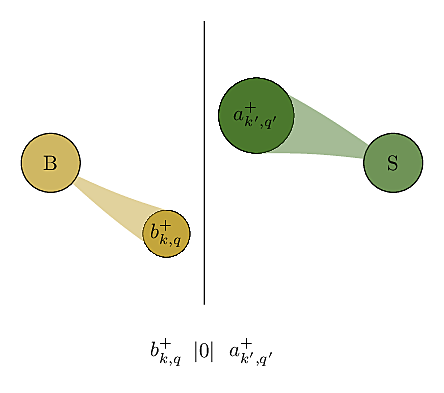}
	\end{subfigure}
	\hfill
	\begin{subfigure}{0.6\textwidth}
		\caption{Order Cancellation}
		\label{subfig:cancellation}
		\centering
		\includegraphics[width=12cm, height=4.6cm, keepaspectratio]{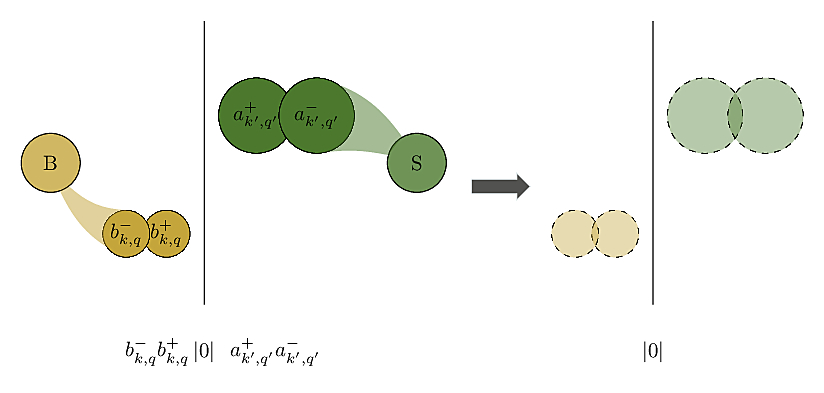}
	\end{subfigure}

	\vspace{1em}

	\begin{subfigure}{0.28\textwidth}
		\caption{Price-Time Priority}
		\label{subfig:priority}
		\centering
		\includegraphics[width=12cm, height=7.3cm, keepaspectratio]{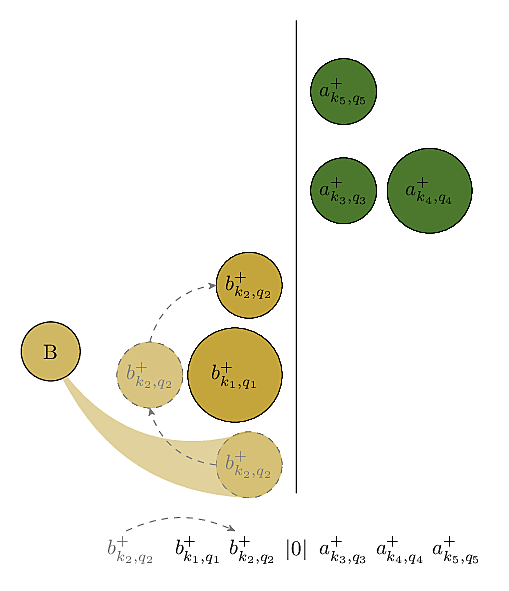}
	\end{subfigure}
	\hfill
	\begin{subfigure}{0.55\textwidth}
		\caption{Order Matching}
		\label{subfig:matching}
		\centering
		\includegraphics[width=12cm, height=7.3cm, keepaspectratio]{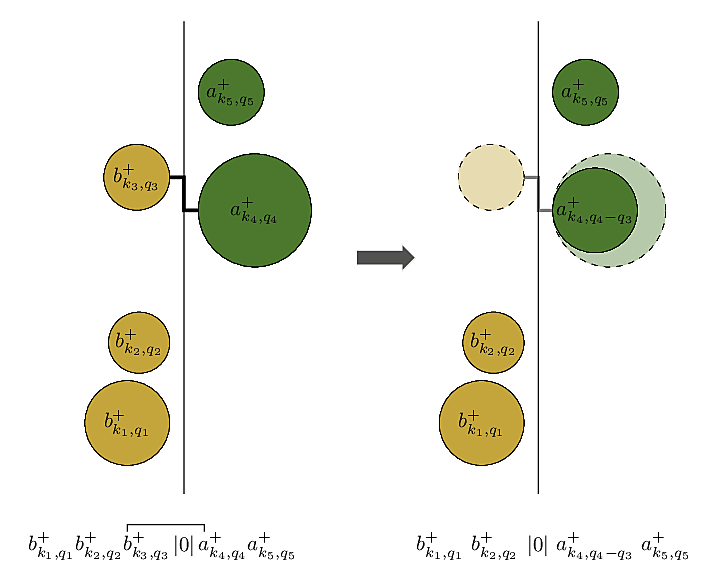}
	\end{subfigure}
\end{figure}

The limit order book (LOB) systematically organizes buy and sell orders based on price and time priorities, governed by stringent rules to ensure market fairness and transparency.
In the following, we describe the rules of order submission, order cancellation, price-time prioritisation, and partial matching of a typical limit order book in terms of creation and annihilation operators in the Dirac notation (\autoref{fig:rules}).
This leads to an algebraic description of all order book states and the associated space of mixed states.

\subsection{LOB Rules and Order Operators}

At any point in time, the state of a limit order book consists of a collection of ask and bid orders that reside in the order book.
Each order is equipped with a designated execution priority based on the time of its submission.
To encode the price-time priorities on the two independent market sides, it is convenient to depict bid and ask orders as operators that act on the order book from opposite sides, left and right, respectively.
For this reason we slightly modify the standard Dirac notation of the vacuum state $\ket{0}$, on which operators conventionally only act from the left, and instead represent the \emph{empty order book} by $|0|$.

\begin{LOB}[Order Submission] \label{rule:creation}
Traders can submit limit orders at a specified price level~$k$ with quantity~$q$.
An incoming ask order is represented by a creation operator $a^+_{k,q}$ that acts on the current order book state from the right, while a bid order is represented by a creation operator $b^+_{k,q}$ that acts from the left.
\end{LOB}

A sequence of successive order submissions captures a history of order book states, for example:
\begin{align*}
	|0|
	\quad \to \quad
	|0| a^+_{k_1,q_1} 
	\quad \to \quad
	b^+_{k_2,q_2} |0| a^+_{k_1,q_1} 
	\quad \to \quad
	b^+_{k_2,q_2} |0| a^+_{k_1,q_1} a^+_{k_3,q_3}
	\quad \to \quad
	\cdots
\end{align*}

The next rule uses the commutator ${[A,B]:= AB - BA}$ and the Kronecker delta~$\delta_{ij}$, satisfying $\delta_{ij}=1$ if $i=j$ and $0$ otherwise.

\begin{LOB}[Order Cancellation] \label{rule:cancellation}
Traders can cancel a previously submitted order.
Ask and bid order cancellations are represented by annihilation operators $a^-_{k,q}$, acting from the left, and $b^-_{k,q}$, acting from the right, that satisfy
\begin{align}
	b^-_{k,q} |0| &= 0 &
	|0| a^-_{k,q} &= 0 \label{eq:DefAnnihilationOperator1} \\
	[b^-_{k,q} , b^+_{j,p}] &= \delta_{kj} \delta_{qp} &
	[a^+_{k,q} , a^-_{j,p}] &= \delta_{kj} \delta_{qp} \label{eq:DefAnnihilationOperator2}
\end{align}
\end{LOB}

The equations in~\eqref{eq:DefAnnihilationOperator1} assert that the probability to observe a cancellation is zero if there is no corresponding order inside the book, while the canonical commutation relations in~\eqref{eq:DefAnnihilationOperator2} encode the fact that annihilation operators only cancel orders of the correct price level and quantity.

Since order book states are the result of successive bid and ask order submissions and cancellations, the list of ask and bid operators are strictly ordered by time.
The next rule specifies how to obtain a price-time ordering from this time-ordered list of operators.

\begin{LOB}[Price-Time Priority] \label{rule:priority}
Orders are organized according to price-time priority.
A list of time-ordered operators is brought into price-time ordering by rearranging the operators using the commutation relations
\begin{align}
	[a^+_{k_1,q_1}, a^+_{k_2,q_2}] 
	= 0 
	= [b^+_{k_1,q_1}, b^+_{k_2,q_2}] \ ,\ 
	k_1\neq k_2 \ ,
\end{align}
such that the price level increases from left to right, while keeping the time-ordering within a group of identical price levels unchanged: 
\begin{align}
	b^+_{k_1,q_1}\ \cdots\  b^+_{k_n,q_n}\ |0|\ a^+_{k_{n+1},q_{n+1}}\ \cdots\ a^+_{k_{n+m},q_{n+m}}\quad ,\ k_{i}\leq k_{i+1} \ .
\end{align}
The price-time priority of a creation operator is encoded by its distance to $|0|$, where operators closer to $|0|$ have higher priority.
\end{LOB}

Finally, to express the execution of a transaction, we represent a pair of matched bid and ask operators by $\wick{\c{b}^+_{k,q} \c{a}^+_{s,p}}$, referred to as 'Wick contraction' in the physics literature.
We also use the Heaviside step function $\theta(q)$, where $\theta(q)= 1$ if $q> 0$, $\theta(q)=\frac{1}{2}$ if $q=0$, and $0$ otherwise.

\begin{LOB}[Order Matching]\label{rule:matching}
Bid and ask orders of highest priority admit a transaction if the bid price is larger or equal to the ask price.
In a transaction order quantities are matched up as far as possible and unmatched quantities remain in the book.
This is represented by the following contraction relation for creation operators:
\begin{align*} \label{eq:matching}
	\wick{\c{b}^+_{k,q} \c{a}^+_{s,p}}
	= \theta(q-p)\  b^+_{k,q-p}
	+ \theta(p-q)\  a^+_{s,p-q} \ ,\
	k \geq s
\end{align*}
where in the case of a perfect match $q=p$ the resulting orders of size $0$ are equivalent to the identity operator.
\end{LOB}

The matching of creation operators corresponds to the following identification of order book states with $k\geq s$:
\begin{align}
\wick{\c{b}^+_{k,q} |0| \c{a}^+_{s,p}} =
\begin{cases}
	|0|a^+_{s,p-q} &  q>p \\
	|0| & q=p \\
	b^+_{k,q-p}|0| & q<p \\
\end{cases}
\end{align}

Let us stress that the price level $k$ of an order $a^+_{k,q}$ is not necessarily the \emph{transaction price} at which the order will be executed.
Instead, the transaction price is determined by the price level of the 'settled order' that is encountered by the 'incoming order'.
See \autoref{sec:Transactions} for a more detailed discussion of transactions.

\subsection{Walking the Book}

In the LOB literature, 'walking the book' refers to an arriving, marketable order that is executed against several orders on the opposite market side.
This notion is nicely reproduced, in a slightly extended form, by the algebra of creation and annihilation operators.

Incoming bid and ask operators act on the order book from the very left or, respectively, right (\autoref{rule:creation}).
An incoming \emph{annihilation} operator 'walks the book' by moving through older creation operators until it either encounters and cancels a matching creation operator or hits the vacuum state (\autoref{rule:cancellation}).
Similarly, an incoming \emph{creation} operator 'walks the book' through older creation orders on the same market side until it reaches its designated position in the price-time ordered queue (\autoref{rule:priority}).
In case the order ends up at the top of the order book, it may encounter orders on the opposite market side, in which case it is matched and executed (\autoref{rule:matching}), effectively continuing its 'walk' through the opposite market side.

\subsection{The Set of Pure Order Book States}
\label{sec:StateSpace}

When all operators have fully walked the book, the state of the order book is uniquely determined by the resulting price-time ordered collection of creation operators
\begin{align*}
	\Big| b^+_{k_1, q_1}\ \ldots\ a^+_{k_n, q_n} \Big\rangle
	:= b^+_{k_1,q_1}\ \cdots\  b^+_{k_j,q_j}\ |0|\ a^+_{k_{j+1},q_{j+1}}\ \cdots\ a^+_{k_{n},q_{n}},
\end{align*}
where $k_i \leq k_{i+1}$ for all $i\in \{1,\ldots, n\}$.
In this list, each operator ${(m_i)}^+_{k_i,q_i}$ is specified by its market side $m_i \in \{a,b\} = \mathcal{M}$, price level $k_i\in \mathcal{K}\subset \mathbb{R}$, and order size $q_i \in \mathcal{Q}\subset \mathbb{R}$.

We call an orderbook configuration that corresponds to $\ket{b^+_{k_1, q_1}\ \ldots\ a^+_{k_n, q_n}}$ a \emph{pure state}.
The set of pure states is given by
\begin{align}
	\mathcal{P} = \biggl\{\ \Big| (m_1)^+_{k_1, q_1} \ldots (m_n)^+_{k_n, q_n} \Big\rangle\  \Big|\ k_i \leq k_{i+1} ,\ n \in \mathbb{N} \biggr\}
	\subset ( \mathcal{M} \times \mathcal{K} \times \mathcal{Q} )^\mathbb{N}
\end{align}
and contains all possible configurations of the limit order book.

Given that the price levels $k_i$, order sizes $q_i$, and the total number of orders $n$ in the order book can become arbitrarily large, the set $\mathcal{P}$ consists of a countably infinite number of pure states $\ket{\psi_k}$, $k \in \mathbb{N}$.
In practice, however, one can always introduce some sufficiently large cutoffs $k_i \leq K$, $q_i \leq Q$, $n\leq N$.
The truncated set of pure states then becomes a large, but finite, set of cardinality at most $(|\mathcal{M}|\times |\mathcal{K}| \times |\mathcal{Q}|)^N$.

\subsection{The Space of Mixed States}

A mixed state is a linear combination $\ket{\Psi} = \sum p_k \ket{\psi_k}$ of several pure states $\ket{\psi_k} \in \mathcal{P}$, where each coefficient $p_k \in [0,1]$ corresponds to the probability that the state $\ket{\psi_k}$ is realized.
In particular, a mixed state is normalized such that $\sum_{\ket{\psi_k}\in \mathcal{P}} p_k = 1$.

It is convenient to also consider unnormalized states.
These are elements of the real vector space $\mathcal{H}$ whose basis is given by the set of pure states $\ket{\psi_k} \in \mathcal{P}$:
\begin{align*}
	\mathcal{H} := \bigg\{ \ket{\Psi} = \sum_{\ket{\psi_k}\in\mathcal{P}} p_k \ket{\psi_k} \quad\bigg|\  p_k\in\mathbb{R}\  \bigg\}  .
\end{align*}
By definition, a mixed state corresponds to a vector $\ket{\Psi} \in \mathcal{H}$ with non-negative coefficients and $L^1$-norm $\norm{ \Psi }_{L^1} = 1$.
Correspondingly, the space of all mixed states is a closed polytope inside of $\mathcal{H}$.

Equipped with the standard $L^2$-product $\mathcal{H}$ becomes a Hilbert space.
It is customary to denote linear functionals on $\mathcal{H}$ by 'bra' vectors $\bra{\Psi}\in\mathcal{H}^\ast$.
In particular, we write $\bra{\psi_k}$ for the dual basis, defined by $\braket{\psi_k | \psi_\ell} = 1$ if $\psi_k = \psi_\ell$ and zero otherwise.

\section{Stochastic Time Evolution of the Limit Order Book}
\label{sec:time_evolution}

Let us now describe how a given initial state of the limit order book $\ket{\psi(t_0)}$ evolves over time.
Throughout this section, we closely follow \cite{Baez2018}, where the general theory of stochastic time evolution of mixed states is laid out in great detail.

\subsection{Markovian Time Evolution}
\label{sec:markov-property}

The limit order book changes exclusively through order submissions or cancellations.
Since any future state must arise from the current state through these \emph{elementary transitions}, the evolution of the LOB state is described by a Markov process.

Order book dynamics should generally be expected to follow a higher-order Markov process.
Market participants seek to maximize the probability of order execution at a favourable price.
To inform their price estimates, individual traders may rely on the recent history of order book configurations, as well as outside information.
Indeed, the position and frequency of incoming orders show time-dependencies that may in part be attributed to a hysteresis in individual trader behaviour.

Instead of attempting to model decisions of individual traders, we here adopt an \emph{effective} description of the dynamical system.
Our main proposal is that, in aggregation, groups of traders can be approximated by a first-order Markov process based on their collective transition rate distribution.

These collective arrival and cancellation rates can be estimated from order flow data.
They are observed to be time dependent, intraday patterns of order flow have been documented for example by \cite{BiaisHS95}.
Even in international Bitcoin markets, in which trading is possible 24/7, \cite{ErossMU19} document activity patterns related to the opening and closing of major markets.
Clustering of transactions in time can be interpreted as the result of short-term increases in arrival and cancellation rates at and near marketable prices.
These are usually modeled using Autoregressive Conditional Duration (ACD) models \cite[see][]{EngleR98,FernandesG06}.

Other variables, like news from outside the order book, may impact the rates of incoming orders also on short time-scales.
Such phenomena may be incorporated in simulations by prescribing an explicit jump-like behaviour of arrival and cancellation rates.

In highly volatile markets, however, the time intervals between events are usually smaller than collective changes in trader sentiment.
In this case quasi-stationary arrival and cancellation rates may serve as good approximations of the evolution of the order book.
Here, we take an indeterministic approach towards the time-dependence of arrival and cancellation rates and rely on estimates from order flow data.

\subsection{The Master Equation}

As a continuous Markov process, the order book satisfies the master equation \cite[cp.][]{vanKampen1992,Weber2017}
\begin{align} \label{eq:MasterEquation}
\frac{\partial}{\partial t} \ket{\Psi(t)} = {H} \ket{\Psi(t)} ,
\end{align}
where the so-called Hamiltonian operator ${H}$ encodes the transition probabilities between order book states.

A formal solution of the master equation is given by the \emph{stochastic time evolution operator} $U(t,t_0) = \exp\big(\int_{t_0}^t H (\tau) d\tau \big)$, which captures the evolution of an initial state $\ket{\Psi(t_0)}$ over time:
\begin{align}\label{eq:Sol_MasterEQ_timedep}
	\ket{\Psi(t)} &= U(t,t_0) \ket{\Psi(t_0)}  \ .
\end{align}
For general Hamiltonians the time evolution operator $U(t,t_0)$ does not admit a closed form expression.
If the Hamiltonian is time-independent, the formula simplifies to ${U (t,t_0) = e^{H (t-t_0)}}$.

As described by \cite{Baez2018}, the Hamiltonian is constructed from \emph{infinitesimal stochastic operators} that represent the elementary transitions of the system:
\begin{align*}
\text{entry of an ask order }\quad E^a_{k,q}&=\left(a^+_{k,q}-1\right) \\
\text{entry of a bid order }\quad E^b_{k,q}&=\left(b^+_{k,q}-1\right)\\
\text{cancellation of an ask order }\quad C^a_{k,q}&=\left(a^-_{k,q}- N_{k,q}^a\right)\\
\text{cancellation of a bid order }\quad C^b_{k,q}&=\left(b^-_{k,q}- N_{k,q}^b\right)
\end{align*}
for some price level $k$ and quantity $q$.
The operators $N_{k,q}^a = a^+_{k,q}a^-_{k,q}$ and $N_{k,q}^b = b^+_{k,q}b^-_{k,q}$ return the number of active bid and ask orders with price level $k$ and quantity $q$ (see \autoref{sec:Observables}).
These need to be included to ensure that the elementary transition operators are infinitesimally stochastic.

The Hamiltonian of the limit order book is necessarily given by
\begin{align}
H = \sum_{k,q}
E^a_{k,q}\alpha_a(k,q)+E^b_{k,q}\alpha_b(k,q)+C^a_{k,q}\omega_a(k,q)+C^b_{k,q}\omega_b(k,q)\ ,
\label{eq:Hamiltonian}
\end{align}
where each transition is weighted by some arrival rate $\alpha$ or cancellation rate $\omega$, respectively.

\subsection{Effective Hamiltonian}

When expressed relative to the best price of the opposite market side, arrival and cancellation rates are observed to be approximately constant, at least on small and intermediate time scales.
Moreover, the distribution of transition rates across prices $k$ and sizes $q$ typically only depend on the current state of the order book through a few macroscopic observables (cf. \autoref{sec:Observables}) -- for example the current best prices and their spread.

Since the distance to the opposite market side and the prevalent spread both depend on the current state of the order book, the arrival rates $\alpha_m(k,q)$ and cancellations rates $\omega_m(k,q)$, $m\in \mathcal{M}=\{a, b\}$, in the Hamiltonian~\eqref{eq:Hamiltonian} must be operators.
This means that when $\alpha_m(k,q)$ or $\omega_m(k,q)$ act on a pure state $\ket{\psi_\ell}$, they return transtition rate distributions that depend on the values of $d$ and $\Delta$ realized in the state $\ket{\psi_\ell}$:
\begin{align*}
	\alpha_m(k,q; \psi_\ell) = \braket{\psi_\ell|\alpha_m(k,q)|\psi_\ell} \\
	\omega_m(k,q; \psi_\ell) = \braket{\psi_\ell|\omega_m(k,q)|\psi_\ell}
\end{align*}
for each market side $m \in \mathcal{M}$, respectively.
Note that the operators $\alpha_m(k,q)$ and $\omega_m(k,q)$ in the Hamiltonian~\eqref{eq:Hamiltonian} are arranged such that they act on the state before a transition occurs.
Thus, the rate of the corresponding transition $E^m_{k,q}$ is determined based on the current state of the order book as required.

Expressions for the transition rate distributions $\alpha_m(k,q; \psi_\ell)$ and $\omega_m(k,q; \psi_\ell)$ can be estimated from order flow data. Usually, the distribution across relative price levels can be described across relative price levels with log-normal distributions or some power-law distribution \cite[see e.g.,][]{BouchaudMP02}.
Order size of incoming orders is found to be varying a lot \cite[see e.g.,][]{GouldPWMFH13}, and often can also be modeled with power-laws or log-normal distributions.

Notably, with the dependence of transition rates on the state of the order book, we obtain a very general non-linear feedback mechanism.
If some key observable of the state of the LOB changes by an event, the effective arrival and cancellation rates react instantanously to this change, leading to a shift in transition probabilities when compared to the previous distribution.

\subsection{LOB as a Compositional System}
As discussed in \autoref{sec:markov-property}, the effective transition rates in the Hamiltonian~\eqref{eq:Hamiltonian} fundamentally arise from the order decisions of individual traders. Our algebraic framework for the limit order book allows for the description of compositions of groups and individuals, each with distinct decision processes and order behaviors. Each individual trader could possess a unique dynamical system that determines their order submission preferences. However, from a systematic perspective, the notion of trader groups is more practical and frequently encountered in the literature. For instance, \cite{FoucaultKK05} categorize traders into institutional and individual traders, while \cite{FoucaultST11} distinguish between patient and impatient traders.

The time evolution of the order book can be described as the result of several groups of traders with distinct order patterns. The impact of their submission and cancellation probabilities is described by individual Hamiltonians $H_g$, where the index $g$ indicates a corresponding group of traders.
The full \emph{effective Hamiltonian} that arises from combining these subsystems is given by $H_\text{eff}=\sum H_g$ which is again of the form~\eqref{eq:Hamiltonian}.
The difference lies in that the rates $\alpha_m(k,q)$ and $\omega_m(k,q)$ are now expressed as sums over population-specific transition rates, reflecting the fundamental model of trader populations.

However, there exists trader-induced clustering or autocorrelation in arrival rates that cannot be ignored\cite[see e.g.,][]{GouldPWMFH13}.
Additionally, general business activity throughout the day induces patterns in the order book \cite[cf.][]{AdmatiP88}.
When submitting orders to the limit order book (LOB), traders often consider the probability that their orders will be executed within a reasonable time frame. There is a trade-off between immediacy and a slightly delayed order execution \cite[see][]{ChoN00}.
The probability of order execution is directly linked to the arrival rates of orders in the LOB.
As shown by \cite{GouldPWMFH13} in their literature review, traders incorporate historical data into their decision processes, influencing when, at which limit price, and in what quantity they submit their orders to the LOB, thereby inducing autocorrelation into arrival and cancellation rates.

The decomposition of the Hamiltonian, as discussed above, allows for an explicit model that accommodates such scenarios. In general, the model does not exclude the concept of autocorrelation in arrival rates. The idea that prior arrival rates influence current rates is also a central notion in the ACD literature mentioned above.

\begin{rem}
At this point, it is also interesting to note that beyond the model presented in this paper, the LOB may be an open Markov process, which can be described by relying on a compositional model framework -- in the sense of \cite{BaezP17} -- and would allow to incorporate trader behavior.
\end{rem}

\section{Simulations}

The algebraic description of the order book above lends itself to a straightforward simulation exercise.
As described in equation~\eqref{eq:Hamiltonian}, the LOB system is driven by the arrivals and cancellations of orders.
With predetermined arrival and cancellation rates, we are able to simulate the time evoultion described by the master equation~\eqref{eq:MasterEquation} using an algorithm due to \cite{Gillespie1977}.

To showcase the method, we simulated two scenarios in a simplified toy model with twenty price levels $\mathcal{K}=\{1, \ldots, 20\}$ and only unit order volumes, i.e. $\mathcal{Q}=\{1\}$.

\subsection{The simulation algorithm}

Starting point of the algorithm described by \cite{Gillespie1977} is the probability that within the next time interval $[t_0, t_0+\delta\tau]$ no transition event occurs.
Using the stochastic time evolution operator~\eqref{eq:Sol_MasterEQ_timedep} with a time-independent Hamiltonian, this probability is given by
\begin{align}
	P_0(\delta\tau) &= \sum_{\ket{\psi_k} \in \mathcal{P}} \bra{\psi_k} \exp(\diag(H) \delta\tau) \ket{\Psi(t_0)} \label{eq:NothingHappens}.
\end{align}
Here $\diag(H)$ denotes the diagonal part of $H$, i.e. the matrix whose non-vanishing entries are given by
\begin{align*}
\diag(H)_{\ell\ell} = \bra{\psi_\ell} H \ket{\psi_\ell} = - \sum_{k,m} \alpha_m(k;\psi_\ell)  - \sum_{k,m} \omega_m(k;\psi_\ell) =: - \lambda(\psi_\ell) .
\end{align*}
$\lambda (\psi_\ell)$ is the decay rate of the state $\ket{\psi_\ell}$, as it is given by the sum of transition rates over all possible events that may occur if the book is in the state $\ket{\psi_\ell}$.

Following \cite{Gillespie1977} and using the expression for $P_0(\delta \tau)$ in~\eqref{eq:NothingHappens}, the probability that a transition away from some initial pure state $\ket{\Psi(t_0)} = \ket{\psi_\ell}$ will occur exactly at time $t_0+\delta\tau$ is given by
\begin{align}
	P(\delta\tau; \psi_\ell) =  \exp(- \lambda(\psi_\ell) \delta\tau)  . 
	\label{eq:SomethingHappens}
\end{align}
Gillespie's algorithm then proceeds by iterating over the following two steps.
First, generate the time $\delta\tau$ at which the next event occurs by drawing from the exponential distribution~\eqref{eq:SomethingHappens} with decay rate $\lambda(\Psi(t_i))$.
Second, determine the new state at $t_{i+1} = t_i+ \delta\tau$ by drawing one transition from all possible ones.
For this, normalize the sum of all arrival and cancellation rates to one by dividing by $\lambda(\Psi(t_i))$ and use the resulting normalized rates as a probability distribution across all possible events.

\subsection{Scenario Parametrization}

To simplify the simulations, we assume that the arrival rates $\alpha_m(k)$ do not depend on the current state $\ket{\Psi(t_i)}$.
The order arrival rates are formed by a single discrete log-normal distribution over a predefined span of relative price levels.
With relative price levels we mean that we associate a rank to an absolute price level in relation to the beginning of the support of the arrival rates.
For the bid price levels the rank or relative price level is formed decreasing, i.e. price level $\rank(k_B = 12)= 1$,  $\rank(k_{B} = 11) = 2$, \ldots, for the ask price levels the rank is determined increasingly, i.e., $\rank(k_A = 9)= 1$,  $\rank(k_{A} = 10) = 2$, \ldots
Bid orders can possibly arrive at price levels $k_B \in \{1,2, \ldots 12\}$ and ask orders $k_A \in \{9,10, \ldots 20\}$.

Following \cite{BiFCK01}, we refer to the discrete, truncated log-normal distribution of arrival rates using the abbreviation DGX (discrete gaussian exponential).
The shape of a DGX is governed by its support and two parameters $\mu$ and $\sigma$.

We also choose the cancellation rates $\omega_m(k)$ such that they explicitley depend on the current state $\ket{\Psi(t_i)}$.
Concretely, we fix the cancellation rate for for each individual ask or bid order in the book to $\omega = 0.1$.
The cancellation rate $\omega_m(k)$ at price level $k$ is then given by the sum over the individual cancellation rates of active orders.
In this way, more orders in the book make it more likely for one to be cancelled in the next step.

In fact, since the cancellation rate on each price level grows with the number of orders residing on this price level, the probability that the next event is a cancellation eventually exceeds the probability of it being another order arrival.
Thus, with this choice we are guaranteed that the system eventually approaches a steady state around which it fluctuates.

We also chose to enforce a constant event rate $\lambda$, i.e. the number of events per time unit is fixed to $\lambda=6$. We achieve this by normalizing arrival and cancellation rates (the latter grow with more orders in the book) to unity with the factor $\frac{\lambda}{\lambda^\ast} = \frac{\lambda}{\sum_{k} \alpha(k) + \omega{k}}$.
Consequently, we expect on average 6 events (arrivals or cancellations) per time unit.

In effect, the distribution around what event happens is entirely driven by the relative weighting of the arrival and cancellation rates.
To make that clear if only two bid order are residing in the book (each having a cancellation rate of 0.1), in our setup it is 5 times more likely that another bid order will arrive in the book and 10 times more likely that either a bid or an ask order arrives than for one of the two residing orders to be cancelled.

In order to show the generality of our setup, we simulated two scenarios.

\begin{figure}
\caption{Arrival Rates}
\label{fig:ArrivalRates}
\centering
\begin{minipage}{0.8\linewidth}
\footnotesize
	The figure depicts the order arrival rate distributions. While in the first scenario, only one single group of traders on each market side is simulated, in the second scenario the parameters for the arrival rates are still symmetric across LOB sides, however, two different groups of traders on each market side are active with two different arrival rate parameterizations and different activity share.
\end{minipage}
\begin{subfigure}{0.49\linewidth}
	\includegraphics[width=\linewidth]{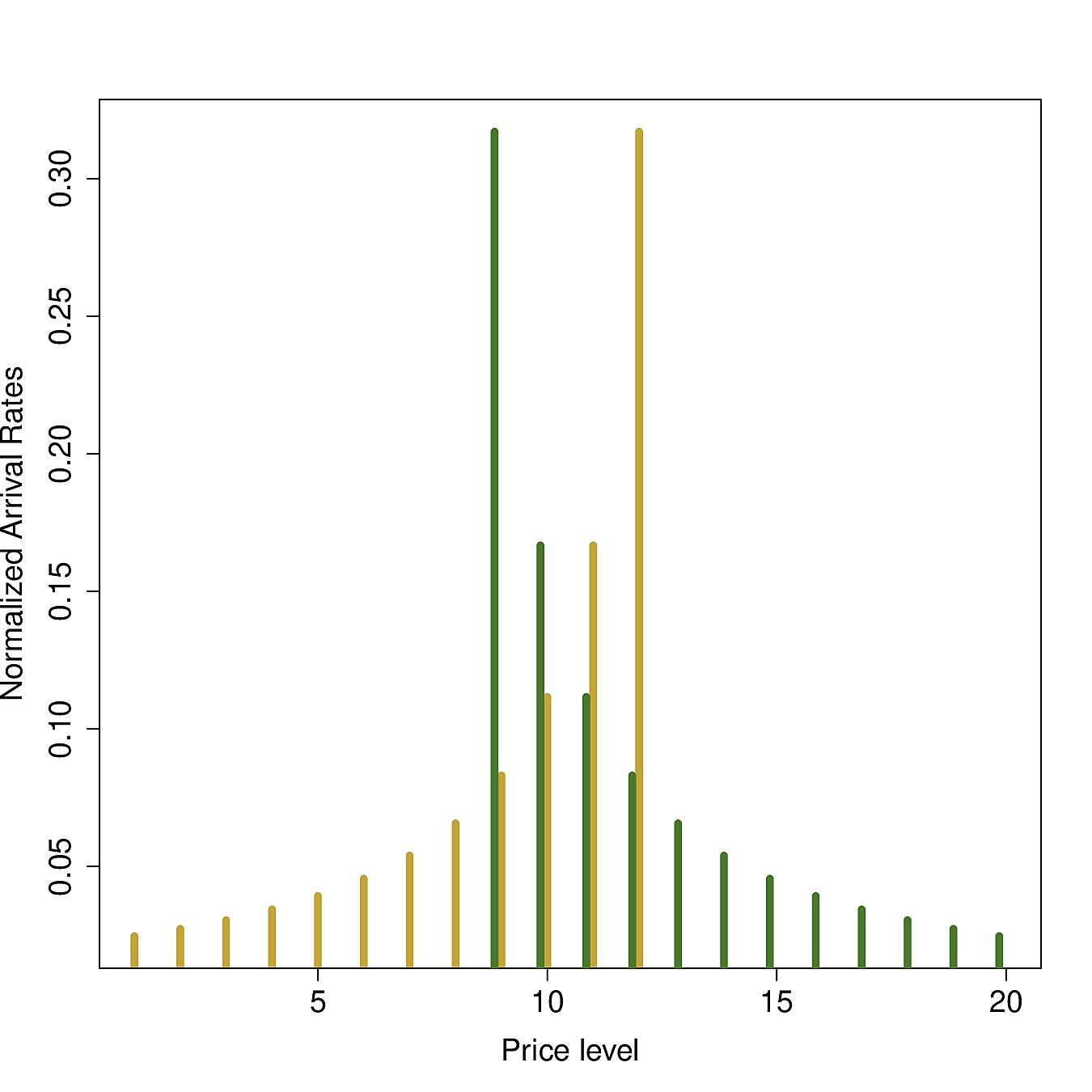}
	\subcaption{Scenario 1} \label{fig:Arrivalrates1g}
\end{subfigure}
\begin{subfigure}{0.49\linewidth}
	\includegraphics[width=\linewidth]{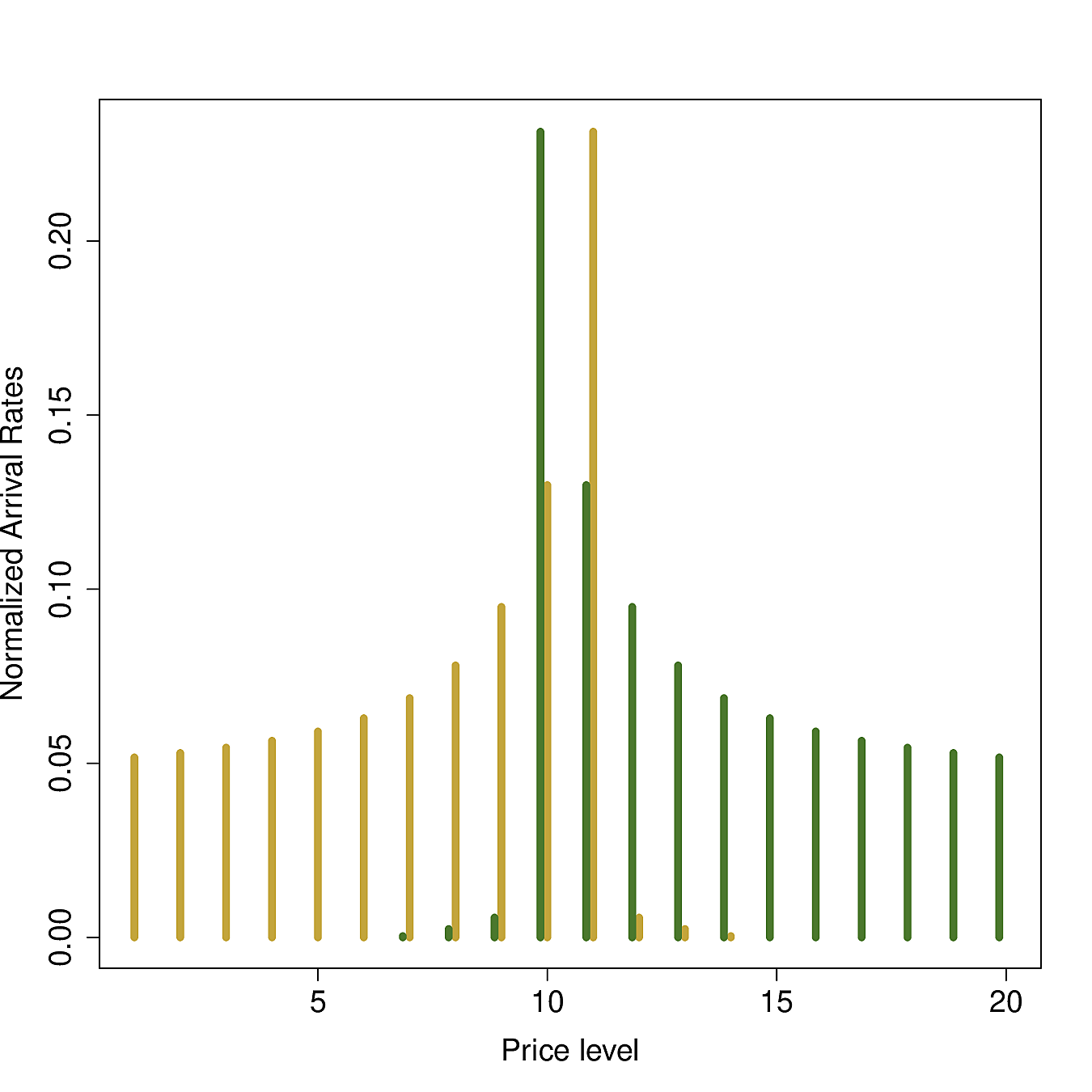}
	\subcaption{Scenario 2} \label{fig:Arrivalrates2g}
\end{subfigure}
\end{figure}

\paragraph{Scenario 1: One group of traders}
First, we use a scenario where there is only a single group of traders on each side of the book.
We assume that the order arrival rates are formed by a DGX with support across 12 price levels.
Hence, in this first scenario, the arrival rates for bid and ask orders will overlap by 4 price levels.
The parameters of this scenario $\mu_1 = 1$ and $\sigma_1=3$ are chosen so that order arrivals are concentrated to the price levels with rank 1 and sharply fall off thereafter.
Figure~\ref{fig:Arrivalrates1g} depicts the arrival rates on both sides of the book.

\paragraph{Scenario 2: Two groups of traders}
In a second scenario, we introduce a second group of traders that submits orders to 14 price levels and whose arrival rates are parametrized with $\mu_2=4$ and $\sigma_2=1$.
Hence, for this group order submissions overlap on 8 price levels.
This implies that in this scenario more arrivals will occur slightly deeper into the book.
Also the dispersion of arrivals across price levels is higher.
We might consider this second group as one in which there more uninformed (higher dispersion) and patient (submissions tend to occur deeper into the book).

Further, we assume that 30\% of all order arrivals are submitted by this second group and the remaining 70\% by the first group from scenario 1. The resulting normalized arrival rate distribution is depicted in Figure~\ref{fig:Arrivalrates2g}.
The arrival rates depicted in Figure~\ref{fig:ArrivalRates} are normalized to 1 for each side. Hence, each side is equally likely to submit an order.

\begin{table}
\caption{Parameterization of the Toy Model Simulations}
\label{tab:sim_parametrization}
\centering
\begin{minipage}{0.9\linewidth}
 \footnotesize
 The table lists the parameters necessary for the simulation. While in the first scenario, only one single group on each market side is simulated, in the second scenario the parameters for the arrival rates are still symmetric across LOB sides, however two different groups on each market side are active with two different arrival rate parameterizations and different activity share.
 \\
\end{minipage}

 \begin{tabular}{ccccccc}
\toprule
&&& \multicolumn{2}{c}{\bfseries Group 1} & \multicolumn{2}{c}{\bfseries Group 2}\\
\cmidrule(lr){4-5} \cmidrule(lr){6-7}
&&	&\itshape Ask & \itshape Bid &  \itshape Ask & \itshape Bid\\[18pt]
\multirow{6}{*}{\bfseries Scenario 1}
& \itshape group share & $s$ & \multicolumn{2}{c}{100\%} &\multicolumn{2}{c}{--}\\[6pt]
		\cmidrule(lr){2-7}
	&\multirow{2}{*}{ \itshape arrival rates}
		&$\mu$ & 1 & 1 & --&--\\
		&&$\sigma$ & 3 & 3 & --&--\\
		\cmidrule(lr){2-7}
	&\itshape cancellation rate &$\omega$ & \multicolumn{2}{c}{0.1}& --&--\\[6pt]
	&\itshape event intensity & $\lambda$ & \multicolumn{2}{c}{6}& --&-- \\[24pt]
\multirow{6}{*}{\bfseries Scenario 2}
	&\itshape group share & $s$ & \multicolumn{2}{c}{70\%} &\multicolumn{2}{c}{30\%}\\[6pt]
		\cmidrule(lr){2-7}
	&\multirow{2}{*}{\itshape arrival rates}
			&$\mu$ & 1 & 1 & 4&4\\
			&&$\sigma$ & 3 & 3 & 1&1\\
	\cmidrule(lr){2-7}
	&\itshape cancellation rate &$\omega$ & \multicolumn{4}{c}{0.1}\\[6pt]
	&\itshape event intensity & $\lambda$ & \multicolumn{4}{c}{6}\\[6pt]

\bottomrule
\end{tabular}
\end{table}

\subsection{Simulation results}

\begin{figure}
\centering
\caption{Simulation Results: Key Observables}
\label{fig:SimKeyObservables}
\begin{minipage}{0.9\linewidth}
\footnotesize
 The figure shows kernel densities of the mean values of key observables based on the observed transaction in 5{,}000 changes of the limit order book simulated in 1{,}000 simulation runs. The values are grouped by the two scenarios. \\
\end{minipage}
\includegraphics[width=\linewidth]{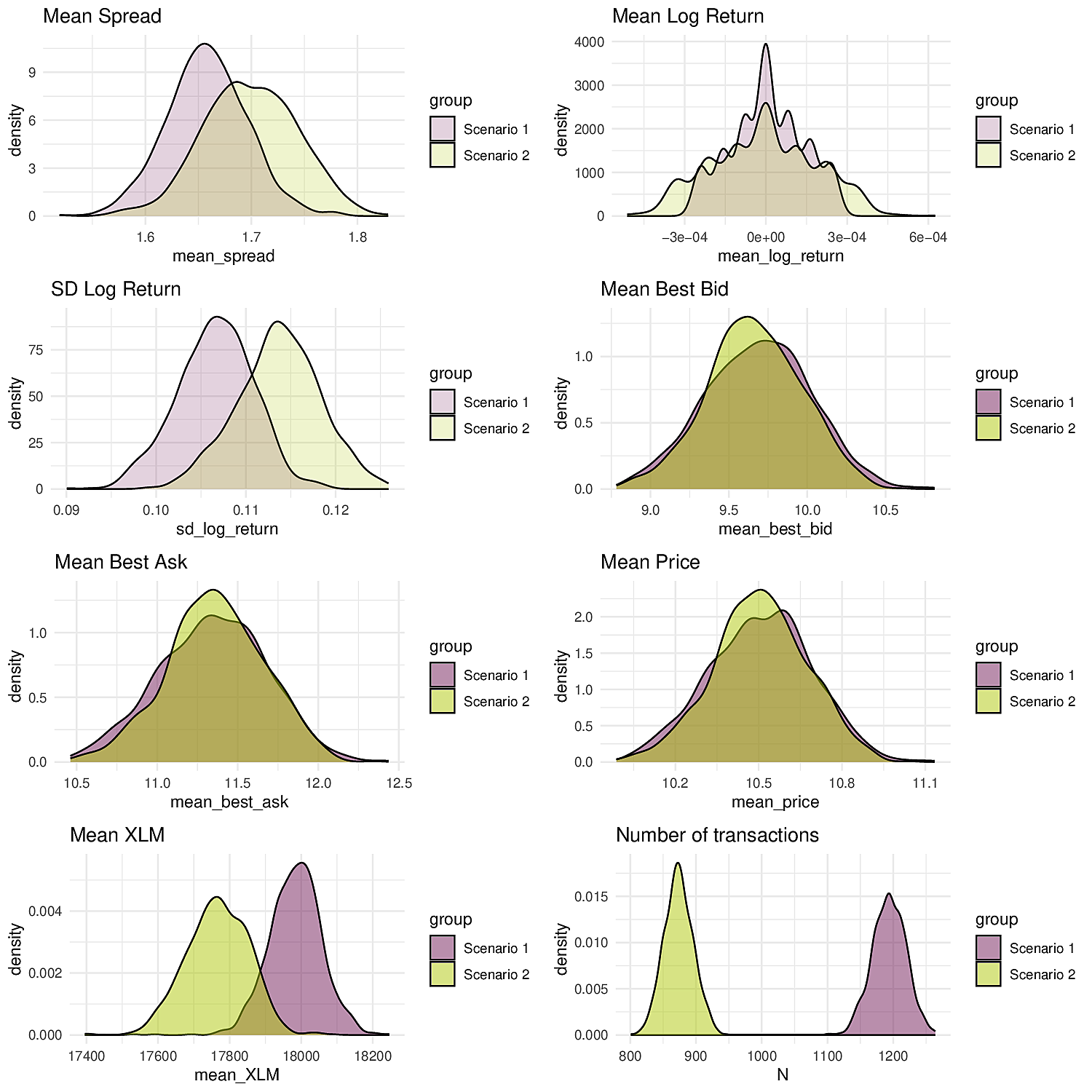}
\end{figure}

\begin{figure}
\centering
\caption{Simulation Results: Order Book and Transactions}
\label{fig:LOBheat}
 \begin{minipage}{0.9\linewidth}
 \footnotesize
  The figure presents the last 100 simulation steps (x-axis) of 50{,}000 simulation runs. The coloring of each price level (y-axis) indicates the volume level of orders residing at each price level. Black dots indicate that a transaction occured on the respective price level. The parameterization of scenario 1 displayed in subfigure~\ref{fig:LOBheat1g} and scenario 2 shown in subfigure~\ref{fig:LOBheat2g} can be found in table~\ref{tab:sim_parametrization}.
 \end{minipage}
\begin{subfigure}{\linewidth}
\includegraphics[width=\linewidth]{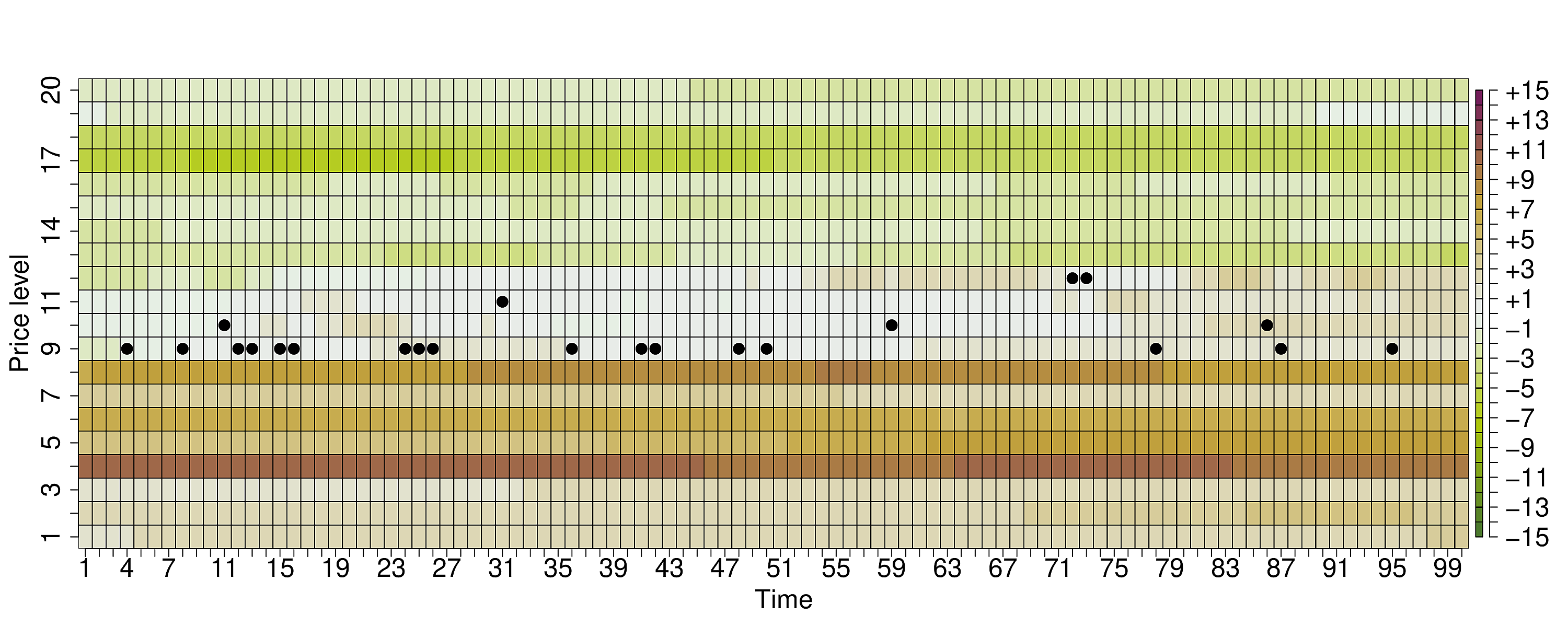}
 \subcaption{Scenario 1} \label{fig:LOBheat1g}
\end{subfigure}
\begin{subfigure}{\linewidth}
\includegraphics[width=\linewidth]{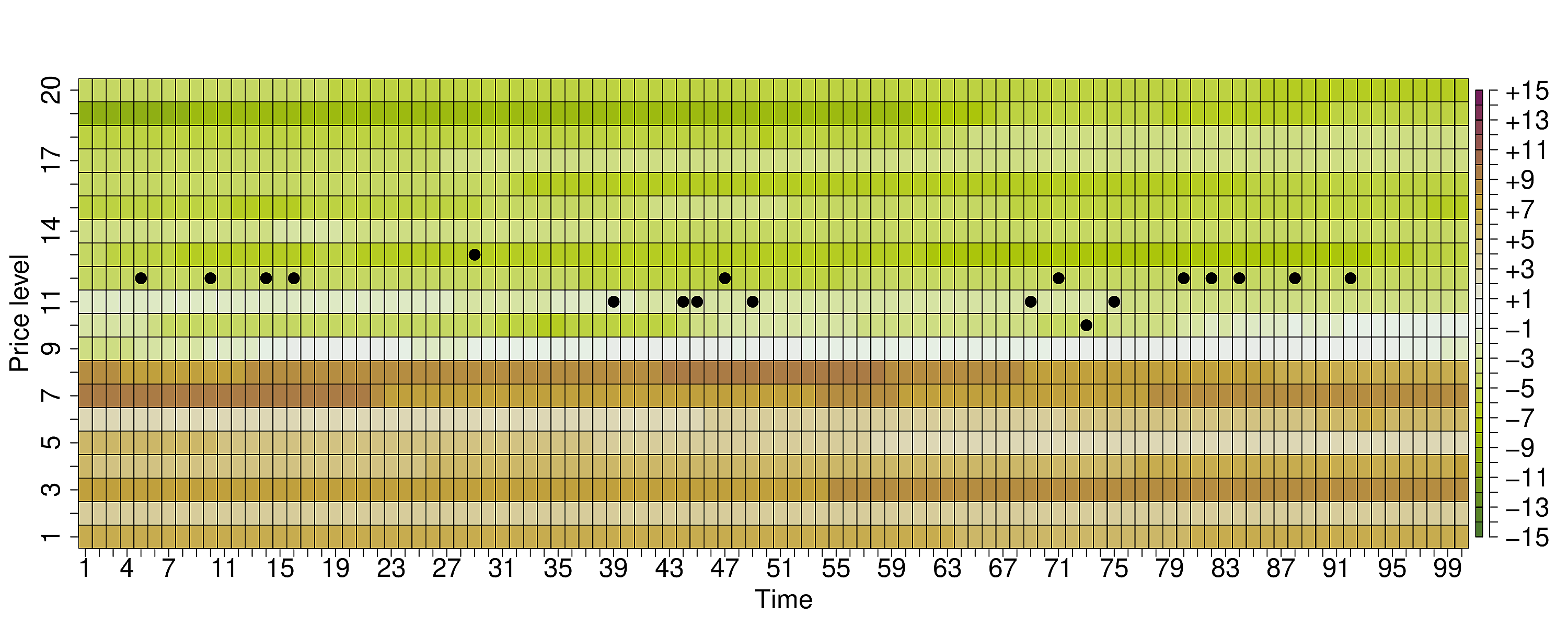}
 \subcaption{Scenario 2} \label{fig:LOBheat2g}
\end{subfigure}
\end{figure}

Figure~\ref{fig:LOBheat} shows the last 100 discrete simulation steps for Scenario 1 as well as for Scenario 2.
Figure~\ref{fig:SimKeyObservables} depicts the distributions of aggregate key observables like the mean of spread, returns, transaction price, best bid price, best ask price, the transaction rate and the XLM liquidity measure (cf. \autoref{sec:Observables}).

The setup of Scenario 1 can be interpreted as a market where there are symetrical trader groups on each market side. Note that the groups may be comprised of traders that differ in their valuation and level of patience.
Economically, since the arrival rates of the two market sides overlap, the valuation of sellers and buyers for the instrument traded is somehow opaque.
In effect, it is unclear whether it is a good deal to sell or buy at some price levels. Such variation in the valuation may occur in reality, if different approaches for the valuation are chosen, different levels of information are prevalent among traders or if they had different preferences. For example, traders placing orders deeper into the book, may hope for a different valuation of the other market side in the future and might just have a low preference for immediacy. So for them hoping to get a better deal in the future, might be option.

In scenario 2, we overlay the arrival rates of two trader groups. The second group has a higher variation and a wider support. We could think of a rather uninformed group in which prices are more randomly chosen from a wider range of price levels. This scenario could be thought of the grouping \cite{FoucaultST11} chose with rather informed institutional and rather uninformed individual traders.

In comparing scenario 1 and 2, we find that the shifted arrival rates (deeper into the book) cause fewer transactions and decrease the transaction rate per unit time.
The spread widens and the return volatility increases in the 2-group scenario.
Therewith, the liquidity as measured by the XLM worsens.

%

\section{Conclusion}
In this article, we have shown that the unconventional framing of the LOB in terms of an operator algebra using the tools usually emploied for the description of stochastic processes in chemestry and physics is useful to describe the dynamics of the LOB.
We highlighted and identified the importance of order arrival and cancellation rates  for the dynamics of the LOB.
The description allows for a compositional setup of the LOB. It is able to connect the order flow ot the different groups of traders and can be made arbitrarily complex.
Even the dynamics of simulataneous trading in  mutiple LOBs for different assets could be incorporated by exapanding the state space the Hamiltonian acts uppon.
This would also necessitate the introduction of arrival and cancellation rates for the other LOBs.
With a simplified toy example, we have shown how the introduction of different groups of traders affects key observables of the LOB.

The simulation is enabled by the formal algebraic describtion of the LOB we introduced which allows for an exact simulation of the LOB with the Gillespie algorithm. Such simulations can be useful to describe differente scenarios, approximate the implications and impact of market design, policy changes, and other environmental factors that affect arrival and cancellation rates.

Future research may use this approach to explore more realistic and more complex simulation settings for the LOB, such as introducing varying oder sizes, adaptive event rates, market side imbalances and more \ldots
These simulations could then be used to estimate latent variables and parameters to observable LOB market data.

\newpage

\onehalfspace
\addcontentsline{toc}{section}{References}
\bibliographystyle{kluwer}
\bibliography{libPhD3}

@Book{Baez2018,
  Title                    = {Quantum Techniques In Stochastic Mechanics},
  Author                   = {Baez, John C. and Biamonte, Jacob D.},
  Publisher                = {World Scientific Publishing Company},
  Year                     = {2018},

  ISBN                     = {9789813226968}
}

@Article{BaezP17,
  Title                    = {A compositional framework for reaction networks},
  Author                   = {Baez, John C. and Pollard, Blake S.},
  Journal                  = {Reviews in Mathematical Physics},
  Year                     = {2017},

  Month                    = {Sep},
  Number                   = {09},
  Pages                    = {1750028},
  Volume                   = {29},

  ISSN                     = {1793-6659},
  Publisher                = {World Scientific Pub Co Pte Lt}
}

@InProceedings{BiFCK01,
  Title                    = {The {'DGX'} distribution for mining massive, skewed data},
  Author                   = {Bi, Zhiqiang and Faloutsos, Christos and Korn, Flip},
  Booktitle                = {Proceedings of the seventh ACM SIGKDD international conference on Knowledge discovery and data mining},
  Year                     = {2001},
  Pages                    = {17--26}
}

@Article{BiaisHS95,
  Title                    = {An Empirical Analysis of the Limit Order Book and the Order Flow in the {Paris} Bourse},
  Author                   = {Bruno Biais and Pierre Hillion and Chester Spatt},
  Journal                  = {Journal of Finance},
  Year                     = {1995},
  Number                   = {5},
  Pages                    = {1655-1689},
  Volume                   = {50}
}

@Article{BouchaudMP02,
  Title                    = {Statistical properties of stock order books: empirical results and models},
  Author                   = { Jean-Philippe Bouchaud and Marc M{\'e}zard and Marc Potters },
  Journal                  = {Quantitative Finance},
  Year                     = {2002},
  Number                   = {4},
  Pages                    = {251-256},
  Volume                   = {2},

  Eprint                   = { 
 https://doi.org/10.1088/1469-7688/2/4/301
 
},
  Publisher                = {Routledge}
}

@Article{ContST10,
  Title                    = {A stochastic model for order book dynamics},
  Author                   = {Cont, Rama and Stoikov, Sasha and Talreja, Rishi},
  Journal                  = {Operations Research},
  Year                     = {2010},
  Number                   = {3},
  Pages                    = {549--563},
  Volume                   = {58},
  Publisher                = {INFORMS}
}

@Article{EngleR98,
  Title                    = {Autoregressive conditional duration: a new model for irregularly spaced transaction data},
  Author                   = {Engle, Robert F and Russell, Jeffrey R},
  Journal                  = {Econometrica},
  Year                     = {1998},
  Number                   = {5},
  Pages                    = {1127--1162},
  Volume                   = {66},

  Publisher                = {JSTOR}
}

@Article{ErossMU19,
  Title                    = {The intraday dynamics of bitcoin},
  Author                   = {Eross, Andrea and McGroarty, Frank and Urquhart, Andrew and Wolfe, Simon},
  Journal                  = {Research in International Business and Finance},
  Year                     = {2019},
  Pages                    = {71--81},
  Volume                   = {49},

  Publisher                = {Elsevier}
}

@Article{FernandesG06,
  Title                    = {A family of autoregressive conditional duration models},
  Author                   = {Fernandes, Marcelo and Grammig, Joachim},
  Journal                  = {Journal of Econometrics},
  Year                     = {2006},
  Number                   = {1},
  Pages                    = {1--23},
  Volume                   = {130},

  Publisher                = {Elsevier}
}

@Article{FoucaultKK05,
  Title                    = {Limit Order Book as a Market for Liquidity},
  Author                   = {Foucault, Thierry and Kadan, Ohad and Kandel, Eugene},
  Journal                  = {Review of Financial Studies},
  Year                     = {2005},

  Month                    = {08},
  Number                   = {4},
  Pages                    = {1171-1217},
  Volume                   = {18},
  Eprint                   = {https://academic.oup.com/rfs/article-pdf/18/4/1171/24421107/hhi029.pdf},
  ISSN                     = {0893-9454}
}

@Article{FoucaultST11,
  Title                    = {Individual investors and volatility},
  Author                   = {Foucault, Thierry and Sraer, David and Thesmar, David J},
  Journal                  = {Journal of Finance},
  Year                     = {2011},
  Number                   = {4},
  Pages                    = {1369--1406},
  Volume                   = {66},

  Publisher                = {Wiley Online Library}
}

@Article{Gillespie1977,
  Title                    = {Exact stochastic simulation of coupled chemical reactions},
  Author                   = {Gillespie, Daniel T},
  Journal                  = {Journal of Physical Chemistry},
  Year                     = {1977},
  Number                   = {25},
  Pages                    = {2340--2361},
  Volume                   = {81},

  __markedentry            = {[jbleher:6]},
  Publisher                = {ACS Publications}
}

@Article{Gomber2002,
  Title                    = {{Der Market Impact: Liquidit{\"a}tsma{\ss} im elektronischen Wertpapierhandel}},
  Author                   = {Gomber, Peter and Schweickert, Uwe},
  Journal                  = {Die Bank},
  Year                     = {2002},
  Number                   = {2002},
  Pages                    = {485--489},
  Volume                   = {7}
}

@Article{GouldPWMFH13,
  Title                    = {Limit order books},
  Author                   = {Gould, Martin D and Porter, Mason A and Williams, Stacy and McDonald, Mark and Fenn, Daniel J and Howison, Sam D},
  Journal                  = {Quantitative Finance},
  Year                     = {2013},
  Number                   = {11},
  Pages                    = {1709--1742},
  Volume                   = {13},

  Publisher                = {Taylor \& Francis}
}

@Book{Harris2003,
  Title                    = {Trading and Exchanges: Market Microstructure for Practitioners},
  Author                   = {Harris, L.},
  Publisher                = {Oxford University Press},
  Year                     = {2003},

  ISBN                     = {9780195144703},
  Lccn                     = {2002004622}
}

@Book{vanKampen1992,
  Title                    = {Stochastic processes in physics and chemistry},
  Author                   = {van Kampen, Nicolaas Godfried},
  Publisher                = {Elsevier},
  Year                     = {1992},
  Volume                   = {1}
}

@Article{Parlour98,
  Title                    = {Price Dynamics in Limit Order Markets},
  Author                   = {Christine A. Parlour},
  Journal                  = {Review of Financial Studies},
  Year                     = {1998},
  Number                   = {4},
  Pages                    = {789-816},
  Volume                   = {11}
}

@Article{Roll1984,
  Title                    = {A simple implicit measure of the effective bid-ask spread in an efficient market},
  Author                   = {Roll, Richard},
  Journal                  = {Journal of Finance},
  Year                     = {1984},
  Number                   = {4},
  Pages                    = {1127--1139},
  Volume                   = {39},

  Publisher                = {Wiley Online Library}
}

@Article{SmithFGK03,
  Title                    = {Statistical theory of the continuous double auction},
  Author                   = {Smith, Eric and Farmer, J Doyne and Gillemot, Laszlo and Krishnamurthy, Supriya and others},
  Journal                  = {Quantitative Finance},
  Year                     = {2003},
  Number                   = {6},
  Pages                    = {481--514},
  Volume                   = {3},

  Publisher                = {Taylor \& Francis}
}

@Article{Weber2017,
  Title                    = {Master equations and the theory of stochastic path integrals},
  Author                   = {Weber, Markus F and Frey, Erwin},
  Journal                  = {Reports on Progress in Physics},
  Year                     = {2017},
  Number                   = {4},
  Pages                    = {046601},
  Volume                   = {80},

  Publisher                = {IOP Publishing}
}

@article{Hoyt72,
 ISSN = {00031305},
 author = {John P. Hoyt},
 journal = {The American Statistician},
 number = {3},
 pages = {45--46},
 publisher = {[American Statistical Association, Taylor & Francis, Ltd.]},
 title = {Generating Functions in Elementary Probability Theory},
 volume = {26},
 year = {1972}
}

@article{Garman1976,
  title={Market Microstructure},
  author={Garman, Mark B},
  journal={Journal of Financial Economics},
  volume={3},
  number={3},
  pages={257--275},
  year={1976},
  publisher={Elsevier}
}

@article{HongPage01,
title = {Problem Solving by Heterogeneous Agents},
journal = {Journal of Economic Theory},
volume = {97},
number = {1},
pages = {123-163},
year = {2001},
issn = {0022-0531},
doi = {https://doi.org/10.1006/jeth.2000.2709},
author = {Lu Hong and Scott E. Page}
}

@article{LilloFM03,
  title={Master curve for price-impact function},
  author={Lillo, Fabrizio and Farmer, J Doyne and Mantegna, Rosario N},
  journal={Nature},
  volume={421},
  number={6919},
  pages={129--130},
  year={2003},
  publisher={Nature Publishing Group UK London}
}

@article{FarmerGLSS04,
author = {J. Doyne Farmer and L\'{a}szl\'{o} Gillemot and Fabrizio Lillo and Szabolcs Mike and Anindya Sen},
title = {What really causes large price changes?},
journal = {Quantitative Finance},
volume = {4},
number = {4},
pages = {383--397},
year = {2004},
publisher = {Routledge},
doi = {10.1080/14697680400008627},
eprint = {
        https://doi.org/10.1080/14697680400008627
}
}

@article{ZarinelliTFL15,
author = {Zarinelli, Elia and Treccani, Michele and Farmer, J. Doyne and Lillo, Fabrizio},
title = {Beyond the Square Root: Evidence for Logarithmic Dependence of Market Impact on Size and Participation Rate},
journal = {Market Microstructure and Liquidity},
volume = {01},
number = {02},
pages = {1550004},
year = {2015},
doi = {10.1142/S2382626615500045},
eprint = {
        https://doi.org/10.1142/S2382626615500045
}
}

@article{CuratoGL17,
author = {Gianbiagio Curato, Jim Gatheral and Fabrizio Lillo},
title = {Optimal execution with non-linear transient market impact},
journal = {Quantitative Finance},
volume = {17},
number = {1},
pages = {41--54},
year = {2017},
publisher = {Routledge},
doi = {10.1080/14697688.2016.1181274},
eprint = {
        https://doi.org/10.1080/14697688.2016.1181274
}
}

@article{TothPLF15,
title = {Why is equity order flow so persistent?},
journal = {Journal of Economic Dynamics and Control},
volume = {51},
pages = {218-239},
year = {2015},
issn = {0165-1889},
doi = {https://doi.org/10.1016/j.jedc.2014.10.007},
author = {Bence Toth and Imon Palit and Fabrizio Lillo and J. Doyne Farmer}
}

@article{TarantoBBLT18a,
author = {Damian Eduardo Taranto and Giacomo Bormetti and Jean-Philippe Bouchaud and Fabrizio Lillo and Bence Toth},
title = {Linear models for the impact of order flow on prices. I. History dependent impact models},
journal = {Quantitative Finance},
volume = {18},
number = {6},
pages = {903--915},
year = {2018},
publisher = {Routledge},
doi = {10.1080/14697688.2017.1395903},
eprint = {
        https://doi.org/10.1080/14697688.2017.1395903
}
}

@article{TarantoBBLT18b,
author = {Damian Eduardo Taranto and Giacomo Bormetti and Jean-Philippe Bouchaud and Fabrizio Lillo and Bence Toth},
title = {Linear models for the impact of order flow on prices. II. The Mixture Transition Distribution model},
journal = {Quantitative Finance},
volume = {18},
number = {6},
pages = {917--931},
year = {2018},
publisher = {Routledge},
doi = {10.1080/14697688.2017.1397283},
eprint = {
        https://doi.org/10.1080/14697688.2017.1397283
}
}

@article{BonartL18,
title = {A continuous and efficient fundamental price on the discrete order book grid},
journal = {Physica A: Statistical Mechanics and its Applications},
volume = {503},
pages = {698-713},
year = {2018},
issn = {0378-4371},
doi = {https://doi.org/10.1016/j.physa.2018.03.002},
author = {Julius Bonart and Fabrizio Lillo}
}

@article{Kyle85,
 ISSN = {00129682, 14680262},
 author = {Albert S. Kyle},
 journal = {Econometrica},
 number = {6},
 pages = {1315--1335},
 publisher = {[Wiley, Econometric Society]},
 title = {Continuous Auctions and Insider Trading},
 volume = {53},
 year = {1985}
}

@article{AdmatiP88,
    author = {Admati, Anat R. and Pfleiderer, Paul},
    title = "{A Theory of Intraday Patterns: Volume and Price Variability}",
    journal = {The Review of Financial Studies},
    volume = {1},
    number = {1},
    pages = {3-40},
    year = {1988},
    month = {01},
    issn = {0893-9454},
    doi = {10.1093/rfs/1.1.3},
    eprint = {https://academic.oup.com/rfs/article-pdf/1/1/3/24434731/010003.pdf},
}

@book{FreyS2009,
  title={The Impact of Iceberg Orders in Limit Order Books},
  author={Frey, S. and Sand{\aa}s, P.},
  series={AFA 2009 San Francisco Meetings Paper},
  year={2009},
  publisher={SSRN}
}

@misc{BleherBleherD2021,
      title={From orders to prices: A stochastic description of the limit order book to forecast intraday returns},
      author={Johannes Bleher and Michael Bleher and Thomas Dimpfl},
      year={2021},
      eprint={2004.11953},
      archivePrefix={arXiv},
      primaryClass={q-fin.TR}
}

@phdthesis{Bleher21,
  title={Essays on the Statistics of Financial Markets},
  author={Bleher, Johannes},
  year={2021},
  school={Eberhard Karls Universit{\"a}t T{\"u}bingen}
}

@article{ChoN00,
author = {Jin-Wan Cho and Edward Nelling},
title = {The Probability of Limit-Order Execution},
journal = {Financial Analysts Journal},
volume = {56},
number = {5},
pages = {28--33},
year = {2000},
publisher = {Routledge},
doi = {10.2469/faj.v56.n5.2387},

eprint = {

        https://doi.org/10.2469/faj.v56.n5.2387



}

}

\begin{appendix}

\section{Appendix}

\subsection{PGF as Polynomial Representation of the LOB Algebra}
\label{sec:pgf-as-polynomial-representation}

The algebra of creation and annihilation operators satisfied by ask and bid orders is well-known in physics.
It admits a standard representation in terms of polynomials in non-commuting formal variables $x_{k,q}$, one for each price level $k$ and each quantity $q$ on the ask side, and analogous formal variables $y_{k,q}$ on the bid side.

This follows from the fact that $\frac{\partial}{\partial x_{k,q}}$ and $x_{s,p}$ satisfy the commutation relations determined by \autoref{rule:cancellation}, while the constant polynomial $1$ provides a vacuum state.
More concretely, $[\frac{\partial}{\partial x_{k,q}} , x_{s,p}] = \delta_{ks} \delta_{qp}$ and $\frac{\partial}{\partial  x_{k,q}} 1 = 0$.

It follows that creation operators can be represented by operators $a^+_{k,q} \equiv x_{k,q}$ and $b^+_{k,q} = y_{k,q}$, acting on a given polynomial by multiplication from the left and right, respectively.
Similarly, $b^-_{k,q} =  \frac{\partial}{\partial y_{k,q}}$ acts as usual differentiation on polynomials, while $a^-_{k,q} = \frac{\partial}{\partial x_{k,q}}$ must be viewed as a slightly unconventional derivative that acts from the right.

This leads to a representation where an order book state is identified with a probability generating function of the form
\begin{align*}
	G( \{x_{k,q}, y_{k,q}\}_{k,q\geq 0} ) = \sum p(n, \{k_i, q_i\}_{i=1,\ldots, n}) \cdot \big(x_{k_1,q_1} \cdots x_{k_i, q_i}  y_{k_{i+1}, q_{i+1}} \cdots y_{k_n, q_n} \big)
\end{align*}
where the sum is over all price-time ordered monomials in $x_{k,q}$ and $y_{k,q}$ of various orders $n$.

\subsection{Observables} \label{sec:Observables}

A specific configuration $\ket{\Psi}$ of the order book contains an enormous amount of information.
Usually, the focus lies on selected descriptive quantities which can be extracted from the order book at any state.
We will call these quantities \emph{observables} and describe them by the action of an operator~$O$ on the order book state $\ket{\Psi}$.
The value of $O$ for a given state $\ket{\Psi}$ can be calculated as
\begin{align*}
    O(\Psi) = \braket{\Psi| O |\Psi}  .
\end{align*}
Given a state $\ket{\Psi}$, the $\nu$-th conditional moment of the observable $O$ is given by\footnote{In quantum mechanics, a similar relation holds, known as the Born rule: $\E[O^\nu;\Psi] = \bra{\Psi} \hat{O}^\nu \ket{\Psi}$. Since we work with stochastic probabilities instead of  quantum mechanical amplitudes, $\bra{\Psi}$ is replaced by a sum over the dual basis $\bra{\psi_k} \in \mathcal{H}^\ast$.}
\begin{align}
	\E[O^\nu;\Psi] = \sum_{\ket{\psi_k} \in \mathcal{P}} \bra{\psi_k} O^\nu \ket{\Psi} . \label{eq:cond_moments}
\end{align}
In combination with the stochastic time evolution operator $U(t,t_0)$, this yields a description for the time evolution of the observable's moments
\begin{align} \label{eq:MomentProjection}
	\E[O^\nu;\Psi(t) ] = \sum_{\ket{\psi_k}} \bra{\psi_k} O^\nu U(t,t_0) \ket{\Psi(t_0)} ,\ t\geq t_0 \ .
\end{align}
Note that the expected value in \Cref{eq:MomentProjection} is conditional on the initial state $\ket{\psi_0}$ at time $t_0$.

Similarly, we can also calculate the expected value of sums and products of distinct operators.
This gives rise to covariance and correlation measures, e.g.,
\begin{align*}
\operatorname{Cov}(O_1,O_2 ; \Psi(t)) = \sum_{\ket{\psi_k} \in \mathcal{P}} \bra{\psi_k} (O_1 -\E[O_1])(O_2-\E(O_2)) \ket{\Psi(t)}.
\end{align*}

In the following, we present a selection of observables that appear throughout the article.

\paragraph{Number and Volume Operators.}
A fundamental observable is the number of active orders on price level $k$ with size $q$.
It is described for the bid and ask side by the corresponding \emph{number operators}
\begin{align*}
	N_{k,q}^a &= a^+_{k,q} a^-_{k,q}, \\
	N_{k,q}^b &= b^+_{k,q} b^-_{k,q}.
\end{align*}
These operators can be utilized to extract several other observables.
In particular, the total number of active orders on price level $k$ of ask or bid type $m\in \mathcal{M} =\{a,b\}$
			\begin{align*}
			 {N}_k^m = \sum_{q} {N}_{k,q}^m  ,
			\end{align*}
the \emph{quantity} of active orders on price level $k$ and the \emph{total quantity} on each market side $m \in\mathcal{M}$
			\begin{align*}
				Q_k^m &= \sum_{q} q N_{k,q}^m  , \\
				Q^m &= \sum_k Q_k^m   ,
			\end{align*}
or the \emph{volume} of active orders at price level $k$ and the \emph{total volume} on each market side
			\begin{align*}
				V_k^m &= k Q_k^m,\\
				V^m &= \sum_k V_k^m  .
			\end{align*}

\paragraph{Best Bid, Best Ask, and Spread.}
There are also operators that extract a global aspect of the configuration of an order book state $\ket{\psi_k}$, e.g. the best bid and best ask prices $\beta_m$, $m\in\mathcal{M}$.
In the following, let the state of the LOB be
\begin{align*}
	\ket{\psi_k} =\  b^+_{k_1,q_1} \  \ldots\ b^+_{k_j,q_j}\ |0|\ a^+_{k_{j+1},q_{j+1}}\ \ldots\ a^+_{k_n,q_n} .
\end{align*}
Then the \emph{best bid} and \emph{best ask} operators act on $\ket{\psi_k}$ as follows:
\begin{align*}
	\beta_b \ket{\psi_k} &= k_j \ket{\psi_k}, \\
	\beta_a \ket{\psi_k} &= k_{j+1} \ket{\psi_k} .
\end{align*}
Note that $k$ on the right hand side is not an operator but the price level associated with the best quote.
Combining the two, one obtains the \emph{spread} operator $\Delta$ and \emph{mid price operator} $\beta_\text{mid}$ as
\begin{align*}
	\Delta &= \beta_a - \beta_b , \\
	\beta_\text{mid} &= \tfrac{1}{2} \big( \beta_b + \beta_a \big)  .
\end{align*}

\paragraph{Liquidity.}
\cite{Harris2003} defines liquidity as "the ability to trade large size quickly, at low cost, when you want to trade."
According to the same source, the notion of liquidity incorporates four dimensions: immediacy of trade execution for a given size, depth, width, and resilience of the market.
Therefore, the spread itself is used frequently as a liquidity measure in the literature.

There are multiple approaches to measure liquidity and we rely on the exchange liquidity measure (${XLM}$) which is based on the concept of implementation shortfall, introduced by \cite{Gomber2002}.
It covers three dimensions of liquidity: depth, width, and immediacy.
The ${XLM}$ (also known as XETRA Liquidity Measure) is composed of liquidity measures for the ask side (${XLM}_a$) and the bid side of the market (${XLM}_b$),
\begin{align}
 {XLM} = {XLM}_a+{XLM}_b, \label{eq:XLM_all}
\end{align}
where
\begin{align}
 {XLM}_a &= 10,000 \frac{\frac{\sum^{\infty}_{k}V^a_k}{\sum_{k} Q^a_k}-\beta_\text{mid}}{
 \frac{\sum^{\infty}_{k}V^a_k}{\sum_{k} Q^a_k}}, \label{eq:XLMA} \\
 {XLM}_b &= 10,000 \frac{\beta_\text{mid}-\frac{\sum^{\infty}_{k}V^b_k}{\sum_{k} Q^b_k}
 }{
 \frac{\sum^{\infty}_{k}V^b_k}{\sum_{k} Q^b_k}
 }.\label{eq:XLMV}
\end{align}

The ${XLM}$ depends on the volume weighted price which can be realized immediately on each side of the market for a round trip order with a certain volume $\bar{V}$, i.e., simultaneously submitting
marketable ask and bid orders with a total volume of $\bar{V}$.
In other words, the ${XLM}$ measures the cost of a round trip order (in basis points).

\subsection{Transactions} \label{sec:Transactions}
We have deferred a detailed discussion of transactions to the appendix since, strictly speaking, their description is not necessary to set up the order book states and dynamics.
Here we first discuss the trading modes of the XETRA order book and explain how one can augment the LOB states to also record information about transactions.
This enables the study of transaction prices and transaction rates, which are not immediately available in the elementary order book states discussed so far.

The XETRA order book is organized as continuous trading augmented by opening-, intraday-, and closing-auctions.
Before stating the rules for these modes, we make a small change in notation:
Instead of the symbol $|0|$ for the empty book, we use $|T_{k,q;t}|$ to notationally record the last price $k$, quantity $q$, and time $t$ at which a transaction occurred.

\renewcommand{\theLOB}{5a}
\begin{LOB}[Continuous Trading]
Assume an incoming order is assigned highest priority and is such that it permits a transaction with its partner on the opposite market side.
Then the orders will be executed at the price of the partner that was already residing in the market and a transaction of the matched-up quantity will be issued at this price.
For an arriving ask order, this results in
\begin{align}
	\big(\ \cdots b^+_{k,q}\ |0|\ \cdots \big)\, a^+_{s,p} &= \cdots \wick{\c{b}^+_{k,q}\ |\, T_{k, \mathrm{min}(q,p);t}\, |\ \c{a}^+_{s,p}} \cdots \ , \label{eq:Auction1}
\end{align}
while for an arriving bid order, we have
\begin{align}
    b^+_{k,q}\, \big(\ \cdots |0|\ a^+_{s,p}\ \cdots \big) &= \cdots \wick{\c{b}^+_{k,q}\ |\, T_{s, \mathrm{min}(q,p);t}\, |\ \c{a}^+_{s,p}} \cdots \ .\label{eq:Auction2}
\end{align}
\end{LOB}

\renewcommand{\theLOB}{5b}
\begin{LOB}[Auction]
Auctions consist of an outcry/call phase, during which incoming orders are collected and ordered by price-time priority as usual, but are \emph{not executed}.
The exchange may provide an indicative pricing to market participants i.e., the price level at which the current order book state would settle if the call phase were to end immediately.
Upon closing of the call phase the transaction price is determined according to the principle of highest traded volume.
Subsequently, orders of highest priority are executed iteratively at the previously determined transaction price.
The transaction is recorded at the transaction price and with the total traded quantity.
\end{LOB}

\begin{rem}
A concrete description of the matching procedure in an auction, like in \Cref{eq:Auction1}~and~\Cref{eq:Auction2} for continuous trading, is possible for concrete situations.
The principle of highest traded volume makes a general formulation exhausting and is not particularly illuminating.
Therefore, we omit a general formulation at this point.
\end{rem}

We can now explicitly introduce the \emph{transaction price}, \emph{transaction quantity}, \emph{transaction volume}, and \emph{transaction time operators}, which extract the corresponding numbers from the last recorded transaction.
E.g. if the current state of the order book is given by 
\begin{align*}
	\ket{\psi_\ell} = z_1 \ldots z_i\, |\, T_{k,q;t}\, |\, z_{i+1} \ldots z_n \ .
\end{align*}
the operators are act as follows
\begin{align*}
	T_K \ket{\psi_\ell}  &= k \ket{\psi_\ell}, \\
	T_Q \ket{\psi_\ell} &= q \ket{\psi_\ell}, \\
	T_V \ket{\psi_\ell} &= kq \ket{\psi_\ell}, \\
	T_t \ket{\psi_\ell} &= t \ket{\psi_\ell}.
\end{align*}
The latter is the basis for an important observable, the \emph{inter-trade duration} $T_{\Delta t} = t_2 -t_1$, which captures the time between two transaction.
Since $T_{\Delta t}$ is non-local in time and operators can only ever extract information from the \emph{current} state they act on, in our current setup $T_{\Delta t}$ can not be represented by an operator.
However, in practice we can easily record the list of all transactions and calculate it from there.

In fact, it would also be possible to modify what we consider to be a fundamental state of the order book in such a way that the current state 'remembers' enough earlier transactions to represent $T_{\Delta t}$ by an operator.
This is related to the fact that every Markov process of order $k$ can be reformulated as a first order Markov process, by considering histories of $k$ subsequent states as the fundamental states of the system.

\end{appendix}

\end{document}